\documentclass[%
  reprint,
  superscriptaddress,
  showpacs,preprintnumbers,
  amsmath,amssymb,
  aps,
  prb,
]{revtex4-1}

\usepackage{graphicx}% Include figure files
\usepackage{dcolumn}% Align table columns on decimal point
\usepackage{bm}% bold math
\usepackage{hyperref}% add hypertext capabilities
\usepackage{color}
\usepackage{ulem}

\usepackage{times}
\usepackage{booktabs}

\newcommand{\hsat}{H_\text{sat}}
\newcommand{\etal}{\textit{et al.}~}
\newcommand{\ham}{\mathcal{H}}
\newcommand{\bigO}[1]{\mathcal{O}\bigl({#1}\bigr)}
\newcommand{\textfrac}[2]{#1/#2}
\newcommand{\vsq}[1]{\left\lvert{#1}\right\rvert_{}^2}
\newcommand{\eps}{\varepsilon}
\newcommand{\norm}[1]{\left\lvert{#1}\right\rvert_{}^2}
\newcommand{\tildeS}{\widetilde{\mathbf{S}}}
\renewcommand{\Re}[1]{\mathrm{Re}\,{#1}}
\renewcommand{\Im}[1]{\mathrm{Im}\,{#1}}
\newcommand{\Qmu}{\mu}

\begin{document}

\preprint{APS/123-QED}

\title{Vortices, skyrmions, and chirality waves in frustrated Mott insulators\\with a quenched periodic array of impurities}% Force line breaks with \\
%\thanks{A footnote to the article title}%

\author{Satoru Hayami}
\affiliation{%
  Department of Physics, Hokkaido University, Sapporo 060-0810, Japan
}

\author{Shi-Zeng Lin}
\affiliation{%
  Theoretical Division, T-4 and CNLS, Los Alamos National Laboratory, Los Alamos, New Mexico 87545, USA
}

\author{Yoshitomo Kamiya}
\affiliation{%
  Condensed Matter Theory Laboratory, RIKEN, Wako, Saitama 351-0198, Japan
}

\author{Cristian D. Batista}
\affiliation{%
  Department of Physics, University of Tennessee, Knoxville, TN 37996, USA
}
\affiliation{% 
Quantum Condensed Matter Division and Shull-Wollan Center, Oak Ridge National Laboratory, Oak Ridge, TN 37831, USA }

\date{\today} 
 
\begin{abstract}
  Finite-${\mathbf Q}$ magnetic instabilities are rather common in frustrated magnets. When the magnetic susceptibility is maximized  at multiple-${\mathbf Q}$ vectors related through lattice symmetry operations, exotic magnetic orderings such as vortex and skyrmion crystals may follow. Here we show that a periodic array of nonmagnetic impurities, which can be realized through charge density wave ordering, leads to a rich phase diagram featuring a plethora of chiral magnetic phases, especially when there is a simple relation between the reciprocal vectors of the impurity superlattice and the magnetic ${\mathbf Q}$-vectors. 
  We also investigate the effect of changing the impurity concentration or disturbing the impurity array with small quenched randomness. Alternative  realizations of impurity superlattices are briefly discussed.
\end{abstract}
\maketitle

\section{Introduction}
\label{sec:Introduction}

The  emergence of nonzero bulk spin-scalar chirality, known as chiral order, has drawn considerable interest in condensed matter physics. Various consequences of a chiral order have been discussed in different  fields ranging from superconductivity to Mott insulators~\cite{Wen_PhysRevB.39.11413,Kawamura_PhysRevLett.68.3785,Momoi_PhysRevLett.79.2081,Bulaevskii_PhysRevB.78.024402}. 
An attractive area of research is generated by potential realizations of chiral liquid states, i.e., states that exhibit chiral order  in absence of  magnetic order~\cite{Kawamura_0953-8984-10-22-004,Kawamura_PhysRevLett.68.3785,Domenge_PhysRevB.72.024433,Hassanieh_PhysRevLett.103.216402}. Another attractive aspect of chiral states is  their potential for inducing  nontrivial topological phenomena, such as topological anomalous Hall effect for electrons coupled to a chiral spin state through the Berry phase mechanism~\cite{berry1984quantal,Aharonov_PhysRev.115.485,Loss_PhysRevB.45.13544,Ye_PhysRevLett.83.3737,Nagaosa_RevModPhys.82.1539,Martin_PhysRevLett.101.156402}. The very large  fictitious magnetic field produced by this mechanism ($10^3$--$10^4$ T) may bring  major advances for spintronics applications~\cite{Zutic_RevModPhys.76.323}. It is then important to understand the physical mechanisms to stabilize the chiral order. 

Noncoplanar magnetic orderings are accompanied by a nonzero local scalar chirality $\langle \chi_{jkl} \rangle  = \langle \mathbf{S}_j \cdot (\mathbf{S}_k \times \mathbf{S}_l) \rangle \neq 0$, where $j$, $k$, and $l$ are three neighboring lattice sites. Recent theoretical studies on frustrated Kondo lattice models have unveiled a general mechanism for stabilizing  noncoplanar magnetic orderings  in itinerant magnets comprising  conduction electrons coupled to localized magnetic moments~\cite{Martin_PhysRevLett.101.156402,Chern:PhysRevLett.105.226403,Kato_PhysRevLett.105.266405,Akagi_PhysRevLett.108.096401,Hayami_PhysRevB.90.060402,Ozawa15,Hayami_PhysRevB.94.024424}. The mechanism relies on the generation of four and higher spin interaction terms, which appear upon expanding in the small Kondo interaction beyond the  Ruderman-Kittel-Kasuya-Yosida (RKKY) level~\cite{Akagi_PhysRevLett.108.096401,Hayami_PhysRevB.90.060402,Ozawa15}.
These multi-spin interactions are relatively weak in strongly coupled  Mott insulators. In terms of a Hubbard model description with hopping amplitude $t$ and on-site Coulomb potential $U$, four spin interactions are of order ${\cal O}(t^4/U^3)$, while two-spin interactions are ${\cal O}(t^2/U)$. 
Consequently, chiral spin textures are less common in Mott insulators with isotropic exchange interactions. Indeed, these systems usually exhibit a conical spiral order with zero net scalar chirality even in an external magnetic field; otherwise either collinear or coplanar orderings. However, recent theoretical studies in both classical~\cite{Okubo_PhysRevLett.108.017206,Rousochatzakis2016,leonov2015multiply,Shizeng_PhysRevB.93.064430,Hayami_PhysRevB.93.184413} and quantum~\cite{Kamiya_PhysRevX.4.011023,Wang_PhysRevLett.115.107201,Marmorini2014} frustrated spin systems, have shown that the interplay between geometric  frustration, thermal~\cite{Okubo_PhysRevLett.108.017206} or quantum fluctuations,~\cite{Kamiya_PhysRevX.4.011023,Wang_PhysRevLett.115.107201,Marmorini2014} magnetic anisotropy~\cite{Rousochatzakis2016,leonov2015multiply,Shizeng_PhysRevB.93.064430,Hayami_PhysRevB.93.184413,Binz_PhysRevLett.96.207202,Binz_PhysRevB.74.214408,Park_PhysRevB.83.184406}, and long-range (dipolar) interactions~\cite{lin1973bubble,malozemoff1979magnetic,kiselev2011chiral}, can stabilize a plethora of multiple-$Q$ spin textures in Mott insulators, some of which have net scalar spin chirality. In this context, it is worth mentioning a recent experimental confirmation of a triple-$Q$ vortex crystal in the scandium thiospinel MnSc$_2$S$_4$ induced by a magnetic field, where geometric frustration and anisotropy seem to play the key role.~\cite{Gao2016Spiral}

In this article we demonstrate that frustrated Mott insulators coupled to a superlattice of nonmagnetic impurities can generate chiral states in the presence of an external magnetic field.  In contrast to Ref.~\onlinecite{Lin_PhysRevLett.116.187202}, where we have shown that a single nonmagnetic impurity nucleates a magnetic vortex over a finite range of magnetic field values above the bulk saturation field $\hsat$, here
we focus on the effects of a {\it periodic array} of non-magnetic impurities {\it below} the saturation field.  A crucial observation is that 
the local saturation field, $\hsat^I$, around an impurity can be higher than $\hsat$ in frustrated magnets with competing ferro- and antiferromagnetic interactions. 
Given that a single impurity nucleates a magnetic vortex around it for $\hsat < H < \hsat^I$, it is natural to ask about the effect of an  array of impurities when $H < \hsat $.
It is known that nonmagnetic impurities tend to reorient the surrounding spins into a less collinear fashion by inducing an effective biquadratic interaction $(\mathbf{S}_j \cdot \mathbf{S}_k)^2$ with a positive (hence antiferroquadrupolar) coupling constant~\cite{Wollny_PhysRevLett.107.137204,Sen_PhysRevB.86.205134,Maryasin_PhysRevLett.111.247201,maryasin2015collective}. 
This rather general observation provides an alternative motivation for studying the magnetic effects of periodic, and nearly periodic, arrays of impurities. 

Motivated by these observations we focus on the low temperature ($T$) physics of a classical spin model. An important prerequisite is to include competing ferro- and antiferromagnetic exchange interactions so that the Fourier transform, $J(\mathbf{q})$,  has multiple degenerate minima, $\mathbf{Q}_1, \mathbf{Q}_2,\cdots$ connected by lattice symmetry transformations.
The ordered array of nonmagnetic impurities may be realized in a hole- or electron-doped system with sufficiently strong off-site Coulomb interactions so that a charge density wave (CDW) order leads to an array of holes or doubly-occupied sites. For instance, certain high-$T_{c}$ superconductors and related materials are known to have a CDW of holes at a hole concentration $x=1/8$ (Refs.~\onlinecite{Tranquada_PhysRevLett.73.1003,tranquada1995evidence}).
Other possible realizations will be discussed later.

The rest of the paper is organized as follows. In Sec.~\ref{sec: model}, we present our model, outline our Monte Carlo (MC) method, and list the observables that we evaluate. In Sec.~\ref{sec: perfect array}, we show the temperature ($T$)-magnetic field ($H$) phase diagram for a perfectly periodic array of nonmagnetic impurities and  a simple relation between the superlattice reciprocal unit vectors and the $\mathbf{Q}$-vectors (a preliminary account of the discussion in Sec.~\ref{sec:a8} was presented in Ref.~\onlinecite{Batista2016}).
A plethora of multi-$\mathbf{Q}$ phases are obtained from our MC simulations. In Sec.~\ref{sec:variational}, we provide an analytical analysis that explains all of the numerically obtained phases close to $T = 0$. 
In Sec.~\ref{sec: more realistic}, we discuss the effects of changing the impurity concentration or introducing small quenched randomness to the impurity lattice.  Sec.~\ref{sec: summary} includes a discussion of potential  realizations of  periodic arrays of nonmagnetic impurities. 

\section{\label{sec: model}
  Model
}
\subsection{Model}

We consider a two-dimensional classical $J_1$-$J_3$ triangular-lattice Heisenberg magnet in a magnetic field with nonmagnetic impurities. In the absence of impurities, the Hamiltonian is 
\begin{align}
  \ham_\text{pure} = J_1\sum_{\langle j,l \rangle} \mathbf{S}_j \cdot \mathbf{S}_l +
  J_3 \sum_{\langle \langle j,l \rangle \rangle} \mathbf{S}_j \cdot \mathbf{S}_l-H\sum_j S^z_j,
  \label{Eq:J1J3} 
\end{align}
with ferromagnetic nearest-neighbor (NN) exchange, $J_1 < 0$, and  antiferromagnetic third NN exchange $J_3 > 0$. $\mathbf{S}_j$ represents a  classical spin located at the site $j$ with $|\mathbf{S}_j|=1$.
The thermodynamic phase diagram of this model Hamiltonian includes a magnetic field induced finite temperature skyrmion crystal phase  for~\cite{Okubo_PhysRevLett.108.017206,leonov2015multiply,Shizeng_PhysRevB.93.064430,Hayami_PhysRevB.93.184413}   
\begin{eqnarray}
 J_3 / \lvert{J_1}\rvert > J^c_3 / \lvert{J_1}\rvert =  1.0256(53).
\label{cond}
\end{eqnarray}
The skyrmion crystal is a state with spontaneously broken chiral symmetry, corresponding to the superposition of harmonic waves with sixfold-degenerate incommensurate wave vectors $\pm\mathbf{Q}_{\nu}$ ($\nu=1$--$3$). These are the wave-vectors that minimize the Fourier transform of the exchange interaction
\begin{eqnarray}
  \label{Eq:Jq}
  J(\mathbf{q})=\sum_{1 \le j \le 3} (J_1 \cos \mathbf{q}\cdot \mathbf{e}_j + J_3 \cos 2 \mathbf{q} \cdot \mathbf{e}_j).
\end{eqnarray}
Here $\mathbf{e}_1 = \mathbf{\hat{x}}$, $\mathbf{e}_2=-\mathbf{\hat{x}}/2+\sqrt{3} \mathbf{\hat{y}}/2$, and $\mathbf{e}_3 = -\mathbf{e}_1 - \mathbf{e}_2$. The incommensurate minima emanate from the $\Gamma$ point (Lifshitz transition) for $J_3 / \lvert{J_1}\rvert > 1/4$ (we will adopt a convention where the lattice spacing is $a = a^{-1} = 1$): 
\begin{align}
  \pm \mathbf{Q}_1 &= \pm Q \mathbf{\hat{x}},
  \notag\\
  \pm \mathbf{Q}_2 &= \pm Q (-\mathbf{\hat{x}}/2+\sqrt{3} \mathbf{\hat{y}}/2),
  \notag\\
  \pm \mathbf{Q}_3 &= \pm Q (-\mathbf{\hat{x}}/2-\sqrt{3} \mathbf{\hat{y}}/2),
\end{align}
with
\begin{align}
  Q = 2 \arccos \left[ \frac{1}{4} \left( 1 + \sqrt{1-\frac{2J_1}{J_3}} \right) \right].
  \label{Eq:Qvalue}
\end{align}
Thus, according to  Eqs.~\eqref{cond} and \eqref{Eq:Qvalue}, the skyrmion crystal phase is only stable for $Q / (2\pi) >  Q_c / (2\pi) = 0.2623(3)$.
For $\lvert{J_1}\rvert/4<$ $J_3 \lesssim J_3^c$, $J(\mathbf{q})$  resembles the bottom of a wine bottle near the $\Gamma$ point, with a weaker $C_6$ anisotropy as the system approaches the Lifshitz transition point $J_3 = |J_1|/4$. In this regime, the phase diagram comprises only the high-temperature paramagnetic phase and the single-${\bf Q}$ conical spiral phase, both of which have no net scalar spin chirality.~\cite{Hayami_PhysRevB.93.184413}

We will first consider the effect of a periodic array of nonmagnetic impurities forming a perfect triangular superlattice. The primitive reciprocal superlattice vectors are
\begin{align}
  \mathbf{K}_{\pm} = \frac{2\pi}{a_\text{imp}} (1,\pm 1/\sqrt{3}),
\end{align}
where $a_\text{imp}$ is the impurity superlattice constant. 
The impurities are introduced in the Hamiltonian by replacing 
\begin{align}
  \mathbf{S}_{j} &\to p_j \mathbf{S}_{j}, 
  \label{Eq:imp}
\end{align}
where $p_j=0$ ($1$) for a nonmagnetic impurity (magnetic) site.
This amounts to introducing the impurity contribution to the Hamiltonian $\ham_\text{imp} = \ham^{J}_\text{imp} + \ham^{H}_\text{imp}$, so that
 \begin{eqnarray}
  \ham = \ham_\text{pure} + \ham^{J}_\text{imp} + \ham^{H}_\text{imp},
\label{hamil} 
\end{eqnarray}
 with
\begin{align}
  \ham^{J}_\text{imp} &= - \sum_j (1 - p_j) \sum_\eta J_{j,\eta} \mathbf{S}_j \cdot \mathbf{S}_{j+\eta},
  \notag\\
  \ham^{H}_\text{imp} &= H \sum_{j}(1-p_j) S_j^z.
  \label{eq:Himp}
\end{align}
Here $\eta$ is the index of the coordination vectors and the different notation $J_{j,\eta} = J_1,J_3$ for the exchange coupling is associated with the $\eta$th coordination vector at each site $j$. 
After presenting phase diagrams for periodic  arrays of impurities (Sec.~\ref{sec: perfect array}), we will introduce a small randomness in the impurity locations  around the superlattice sites (Sec.~\ref{subsec:Skyrmion state by slight charge fluctuations}).

\subsection{Monte Carlo method}
We perform classical MC simulations of  our spin Hamiltonian $\ham$ given in Eq.~\eqref{hamil} for several impurity configurations to be specified below. Our simulations are carried out with the standard Metropolis local updates supplemented with the over-relaxation method.~\cite{Brown1987} The lattice has $N=L^2$ sites (including impurity sites) with $L = 48$, 64, 72, 80, and 96 and we impose the periodic boundary conditions in both directions. 
We first perform simulated annealing for $10^5$--$10^6$ MC sweeps (MCS) to find  low energy configuration, which is then followed by equilibration steps and a subsequent sampling process of $10^5$--$10^7$ MCS at the target temperature. The statistical errors are estimated from $5$--$64$ independent runs. 

We calculate the specific heat, $C$, the uniform magnetic susceptibility $\mathrm{d}M/\mathrm{d}H$, and the spin and the chirality structure factors. The spin structure factor $S_{s}^{\alpha\alpha}(\mathbf{q})$ ($\alpha=x,y,z$) is given  by
\begin{align}
\label{Eq:spinstructure}
S_{s}^{\alpha\alpha}(\mathbf{q}) &=\frac{1}{N} \sum_{j,l} \langle S_j^{\alpha} S_{l}^{\alpha}  \rangle e^{i \mathbf{q}\cdot(\mathbf{r}_j-\mathbf{r}_l)},
\notag\\
S_{s}^{\perp}(\mathbf{q}) &= S_{s}^{xx}(\mathbf{q}) + S_{s}^{yy}(\mathbf{q}).
\end{align}
We define the chirality structure factor $S_{\chi}^{\mu}(\mathbf{q})$ for the upward and downward triangles ($\mu = u,d$, respectively) as
\begin{eqnarray}
  \label{Eq:chiralstructure}
  S_{\chi}^{\mu}(\mathbf{q})=\frac{1}{N} \sum_{\mathbf{R},\mathbf{R}' \in \mu} \langle \chi^{}_{\mathbf{R}} \chi^{}_{\mathbf{R}'} \rangle e^{i \mathbf{q}\cdot(\mathbf{R} - \mathbf{R}')}, 
\end{eqnarray}
where $\mathbf{R},\mathbf{R}'$ run over sites of the specified sublattice, $\mu = u, d$, of the dual honeycomb lattice.
$\chi^{}_{\mathbf{R}} =\mathbf{S}_j \cdot \left(\mathbf{S}_k \times \mathbf{S}_l\right)$ is the spin scalar chirality associated with a triangle centered at $\mathbf{R}$,
where $j,k,l$ are the sites aligned counterclockwise on the triangle. We also introduce the following notations for the total scalar chirality associated with the up ($\chi^u$) and the down ($\chi^d$) triangles,
\begin{align}
  \chi^{\mu} &= \frac{1}{N}\sum_{\mathbf{R}\in \mu} \chi^{}_{\mathbf{R}}~~\text{for $\mu=u,d$},
\end{align}
and their sum,
\begin{align}
  \chi^\text{tot} &= \chi^{u} + \chi^{d},
\end{align}
which is the total scalar chirality.

\section{Periodic array of impurities}
\label{sec: perfect array}

\begin{figure}[b]
\begin{center}
  \includegraphics[width=0.52 \hsize]{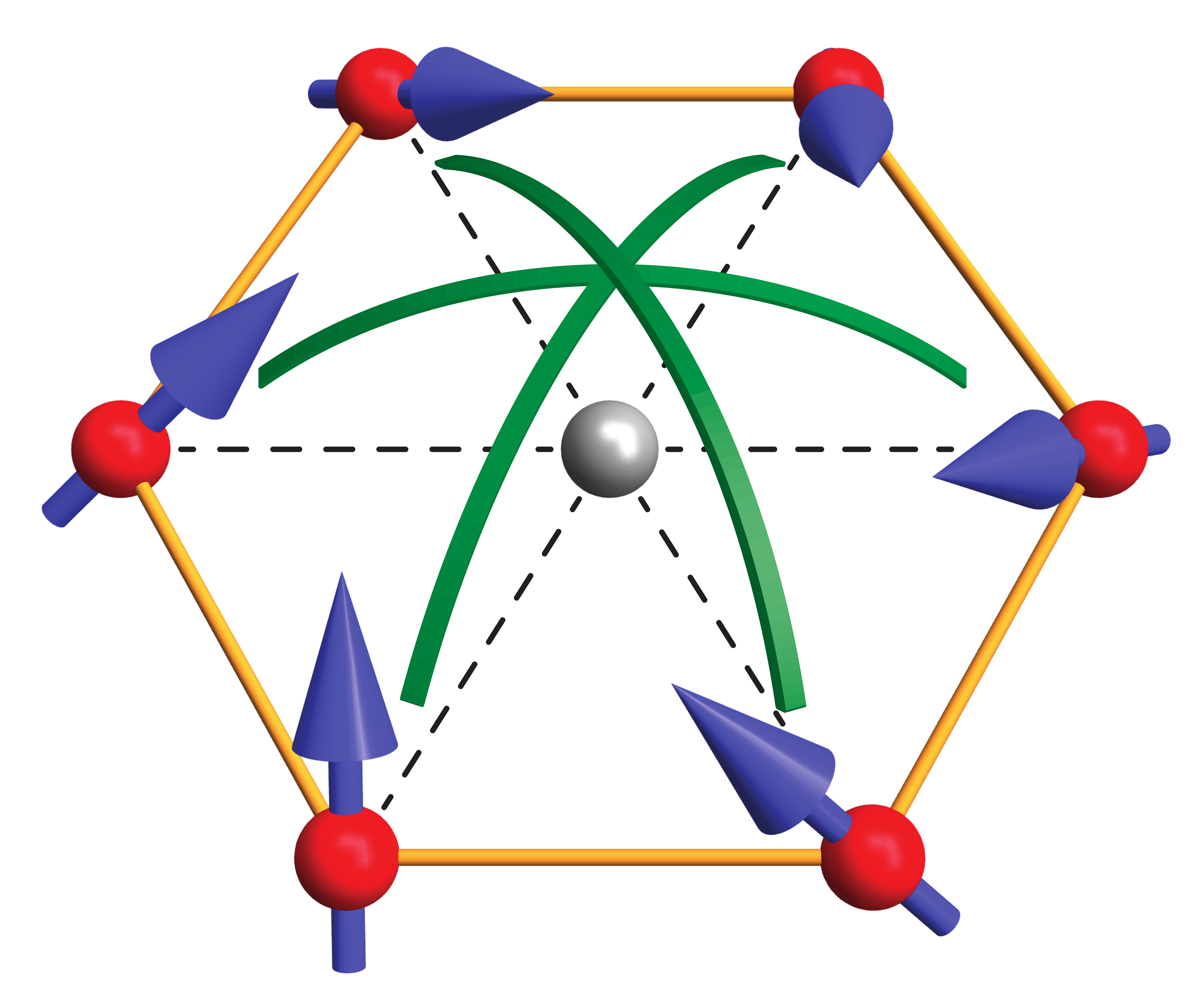}
\end{center}  
\caption{
  \label{Fig:vortex}
  Schematic picture for a local configuration in a magnetic field around a nonmagnetic impurity (at the center). The straight solid (dashed) lines represent the remaining (removed) ferromagnetic exchange $J_1$ interactions. The arc lines represent the antiferromagnetic exchange $J_3$.
}
\end{figure}

\begin{table*}[htb!]
\begin{center}
  \caption{
    Classification of the ordered phases in the classical $J_1$-$J_3$ Heisenberg model with periodic nonmagnetic impurities. Here  $J_3 / \lvert{J_1}\rvert \approx 0.854$ and the period of the impurity superlattice is  $a_\text{imp} = 8$.
    From the fourth to the sixth column, our notation of ``$n+n_\text{sub} (+n'_\text{sub})$'' means that the corresponding structure factor has $n$ dominant peaks as well as $n_\text{sub}$ subdominant (and $n'_\text{sub}$ even smaller) peaks. 
  }
  \vspace{8pt}
  \scalebox{0.75}{
    \begin{tabular}{lcccccc}
    \toprule
    \hline\hline
    phase &
    \parbox[t]{65pt}{nonzero net \\scalar chirality} &
    \parbox[t]{70pt}{chirality of up and down triangles} &
    \parbox[t]{90pt}{number of (quasi-)Bragg peaks in $S_s^{\perp}(\mathbf{Q})$} &
    \parbox[t]{90pt}{number of (quasi-)Bragg peaks in $S_s^{zz}(\mathbf{Q})$\\at $\mathbf{q} \ne 0$} &
    \parbox[t]{90pt}{number of Bragg peaks in $S_\chi^{u/d}(\mathbf{Q})$ at $\mathbf{q} \ne 0$} &
    \parbox[t]{90pt}{broken point-group symmetry (of the lattice with impurities)}\\
    \midrule
    \hline
     ferrochiral 3$Q^M$-6$Q^\chi$ vortex crystal & \checkmark & $\chi^u = \chi^d$ & 3 & 0 & 3+3 & --- \\
    ferrochiral 3$Q^M$ vortex crystal & \checkmark & $\chi^u = \chi^d$ & 3 & 3 & 0 & ---\\
    ferrochiral 3$Q^M$-1$Q^\chi$ spiral & \checkmark & $\chi^u = \chi^d$ & 2+1 & 1 & 1 & $C_6$\\
    ferrichiral 3$Q^M$-2$Q^\chi$ spiral I & \checkmark & $\lvert{\chi^u}\rvert \ne \lvert{\chi^d}\rvert$ & 1+2 & 2+1 & 2+1 & $C_6$\\
    ferrichiral 3$Q^M$-2$Q^\chi$ spiral II & \checkmark & $\lvert{\chi^u}\rvert \ne \lvert{\chi^d}\rvert$ & 1+1+1 & 1+1+1 & 1+1+1 & $C_6$\\
   antiferrochiral 1$Q^M$ spiral & No & $\chi^u = -\chi^d$ & 1 & 1 & 0 & $C_6$ \\
     ferrochiral 3$Q^M$-2$Q^\chi$ spiral & \checkmark & $\chi^u = \chi^d$ & 1+2 & 1 & 2+1 & $C_6$ \\
    vertical 1$Q^M$ spiral & No & --- & 1 & 1 & 0 & $C_6$ ($C_3$ for $H = 0$)\\
    \hline \hline
    \bottomrule
    \vspace{2pt}
  \end{tabular}
  }
\label{tab_a8}
\end{center}
\end{table*}

\begin{figure*}[t!]
\begin{center}
  \includegraphics[width=1.0 \hsize,bb=0 0 1413 521]{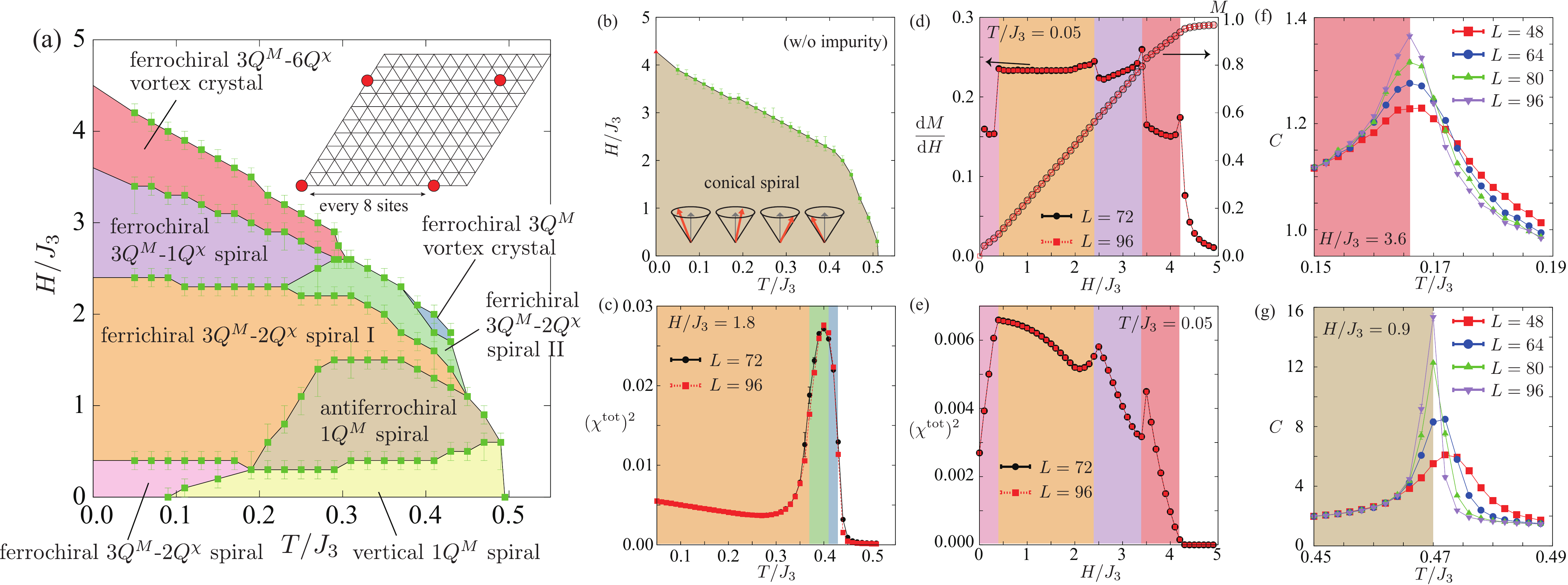}
\end{center}  
\caption{
  \label{Fig:phase-diagram_a8}
  (a) Phase diagram for a perfect periodic array of impurities ($a_\text{imp} = 8$) with $J_3 / |J_1| \approx 0.854$ ($Q = 2\pi/4$). The inset shows the impurity superlattice. 
  (b) Phase diagram of $\ham_{\text{pure}}$.
  (c) $T$-dependence of the net scalar chirality at $H/J_3=0.18$, 
  (d) magnetization curve and its $H$-derivative at $T/J_3=0.05$, and 
  (e) $H$-dependence of $(\chi^\text{tot})^2$ at $T/J_3=0.05$, for the model including a  periodic array of impurities. 
   (f), (g) $T$-dependence of the specific heat at (f) $H/J_3=3.6$ and (g) $H/J_3=0.9$. 
}
\end{figure*}

We start our discussion with the case where magnetic impurities form a perfect triangular superlattice. We first require commensurability between the superlattice reciprocal vectors, $\mathbf{K}_\pm$, and $\pm\mathbf{Q}_{1 \le \nu \le 3}$, which is expected to enhance the constructive interference between the impurity superlattice and the spin texture (we will relax this condition later). For the sake of concreteness, we fix $J_3 / \lvert{J_1}\rvert = 1/(4 - 2\sqrt{2}) \approx 0.854$ unless otherwise specified, which corresponds to $Q / (2\pi) = 1/4$ according to Eq.~\eqref{Eq:Qvalue}. Given that  $J_3<J_3^c$, the phase diagram in absence of impurities only includes the single-${\boldmath Q}$ conical spiral state  shown in Fig.~\ref{Fig:phase-diagram_a8}(b). Such a state is not chiral because $\chi^u$ and $\chi^d$ cancel each other. 
For nonzero $H$, this quasi-long-range ordered state completely breaks the $C_6$ symmetry of $\ham$ because of the associated  bond ordering, whereas the $C_6$ symmetry is broken down to $C_3$ for $H = 0$ as there is a continuous symmetry operation that can change the sign of  the vector chirality. In both cases, our results are consistent with the single-step first-order transition found for the zero-field $J_1$-$J_3$ model with classical XY and Heisenberg spins in Refs.~\onlinecite{Tamura2008,tamura2011first,tamura2014phase}, where the symmetry is O(2) and O(3), respectively, corresponding to the cases with and without the magnetic field in the present consideration. 

As described in the introduction, the nonmagnetic impurities in a frustrated magnet with competing ferro- and antiferromagnetic exchanges make the local saturation field for the surrounding spins larger than the bulk value ($\hsat^I > \hsat$). This is so because the nearest neighbor spins of the non-magnetic impurity feel a molecular field, parallel to the applied field, which is lower than the molecular field acting on other spins. This effect is of course present for any value of the external field: the spins that surround a non-magnetic impurity have a lower magnetic susceptibility because the 
impurity removes the ferromagnetic ($J_1$) bonds connecting them to  the missing spin (Fig.~\ref{Fig:vortex}). Because the field inducing the $z$ spin component is smaller than the average for these spins, their $xy$ component becomes larger at low energies. Moreover, we can anticipate that the resulting local spin configuration near each impurity is likely to be a vortex as in the case with $\hsat < H < \hsat^I$,~\cite{Lin_PhysRevLett.116.187202} because of the competition between $J_1$ and $J_3$ for the six spins surrounding the impurity (Fig.~\ref{Fig:vortex}). The rest of the spins have to accommodate their configuration to the local ``boundary condition" imposed by each impurity. 

Below, we consider two representative cases where the superlattice constant for the periodic impurities is $a_\text{imp} = 8$ and $a_\text{imp} = 4$. They correspond to simple relations between the superlattice reciprocal vectors and the ${\boldmath Q}$ vectors, namely, $\mathbf{K}_+ + \mathbf{K}_- = \mathbf{Q}_1$ and $\mathbf{K}_+ + \mathbf{K}_- =  2\mathbf{Q}_1$, respectively.
We will demonstrate that such a commensurate impurity superlattice  produces a drastic change of the magnetic phase diagram.
Our finite-size scaling analysis to characterize the phases is summarized in Appendix~\ref{sec:Finite-size scaling of each phase}.

\begin{figure*}[p!]
\begin{center}
  \includegraphics[width=0.85\hsize, bb=0 0 740 1010]{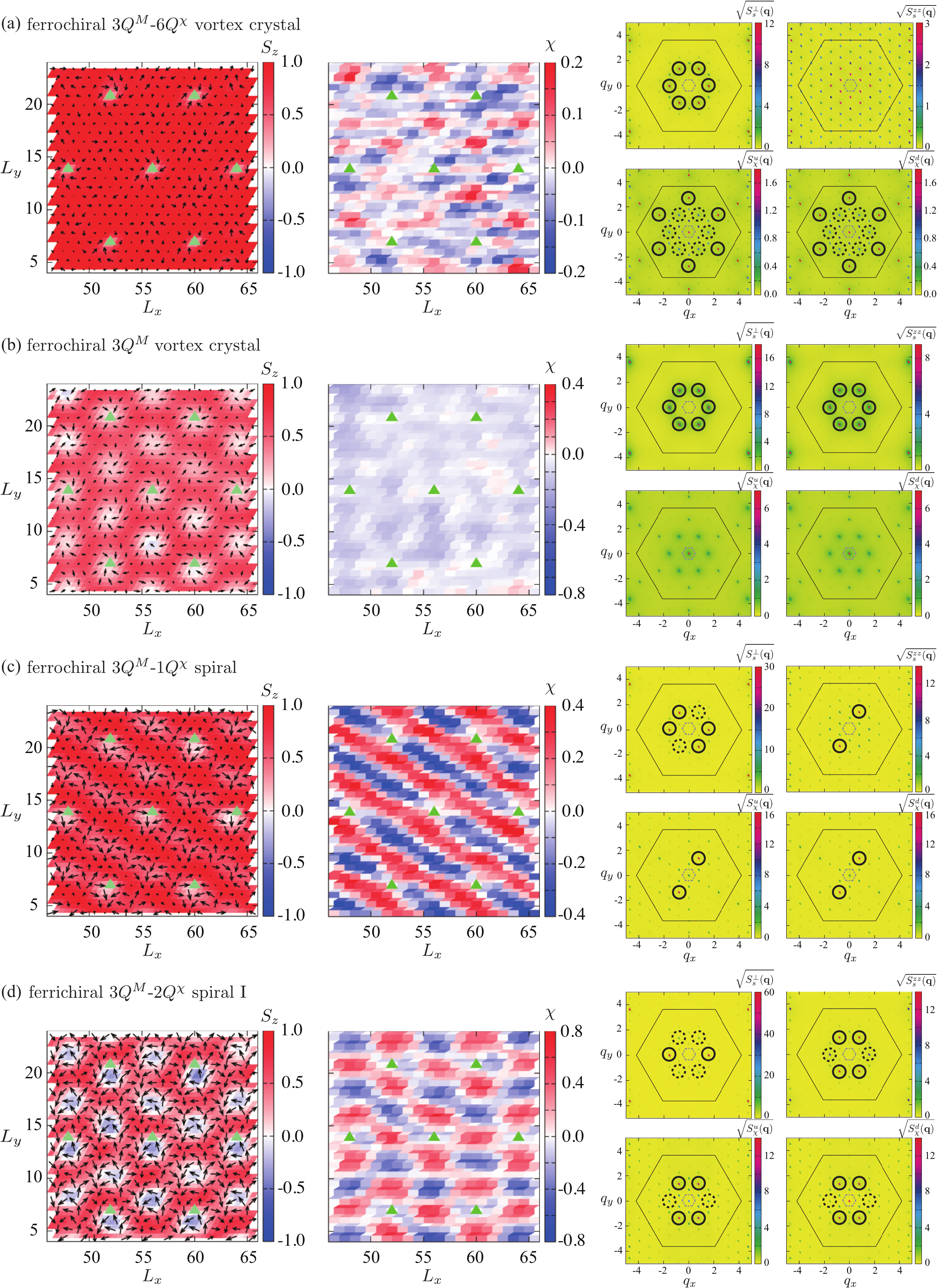}
\end{center}  
\caption{
  \label{Fig:phase1-4_a8}
  Snapshots of the spin (left) and the chirality (middle) configurations in the ordered phases for $J_3 / \lvert{J_1}\rvert \approx 0.854$ ($Q = 2\pi/4$) when periodic impurities (triangles) with $a_\text{imp} = 8$ are present. The spin and the chirality structure factors are shown on the right: (a) ferrochiral 3$Q^M$-6$Q^\chi$ vortex crystal ($H/J_3=4.0$ and $T/J_3=0.05$), (b) ferrochiral 3$Q^M$ vortex crystal ($H/J_3=1.9$ and $T/J_3=0.41$), (c) ferrochiral 3$Q^M$-1$Q^\chi$ spiral ($H/J_3=3.0$ and $T/J_3=0.05$), and (d) ferrichiral 3$Q^M$-2$Q^\chi$ spiral I ($H/J_3=1.5$ and $T/J_3=0.05$). The circles with solid (dashed) lines indicate the dominant (subdominant) $\mathbf{q} \ne 0$ peak(s).
  Note that the $\mathbf{q}=0$ component is removed from $S_s^{zz}(\mathbf{q})$.
  The hexagon with a solid (dashed) line shows the first Brillouin zone (of the impurity superlattice).
  To obtain the smooth spin configurations, we integrate out short wavelength fluctuations by averaging over 50-500 MCS.
}
\end{figure*}

\begin{figure*}[p!]
\begin{center}
  \includegraphics[width=0.85\hsize, bb=0 0 740 1010]{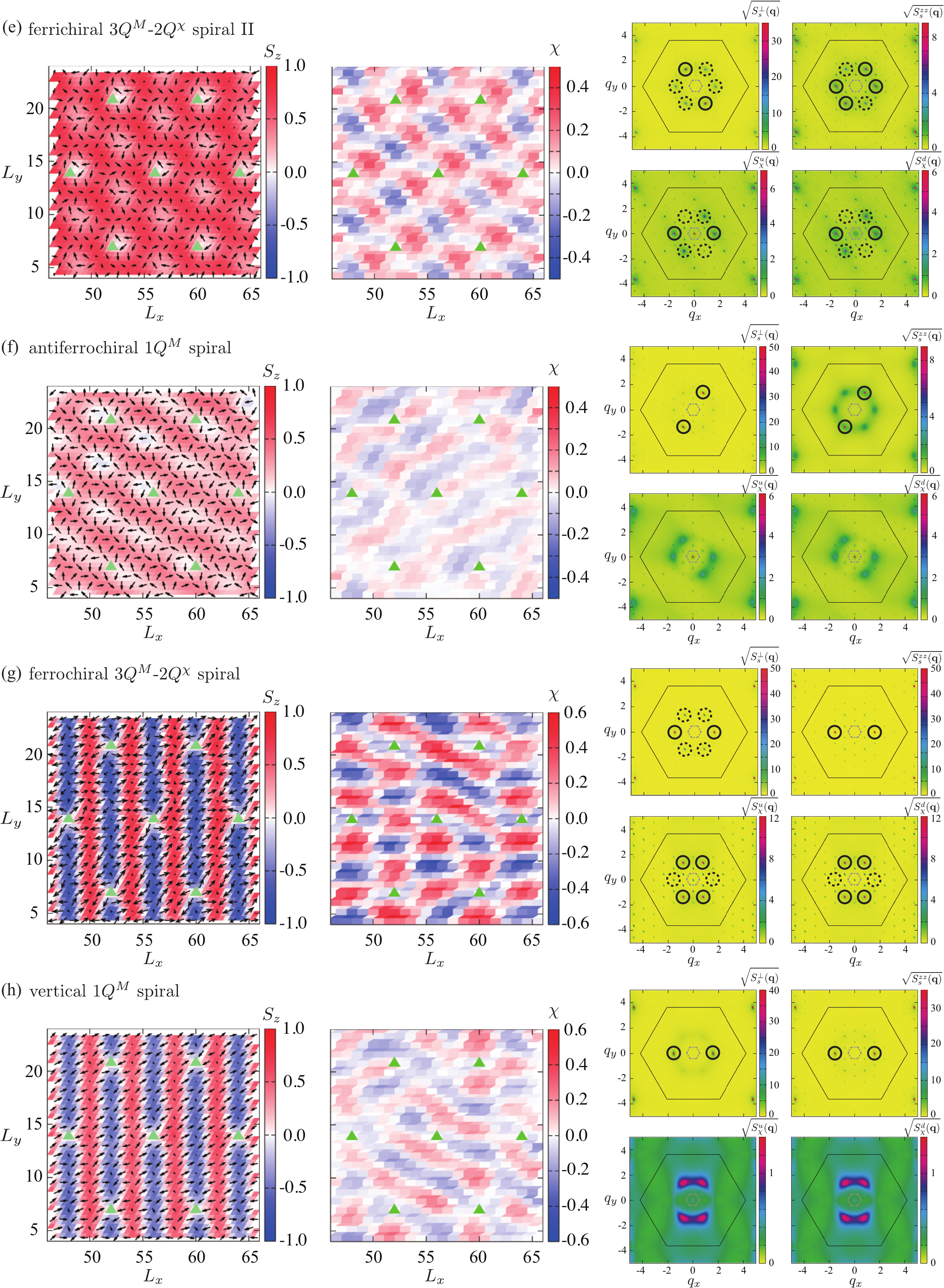}
\end{center}  
\caption{
  \label{Fig:phase5-8_a8}
  Snapshots of the spin (left) and the chirality (middle) configurations in the ordered phases for $J_3 / \lvert{J_1}\rvert \approx 0.854$ ($Q = 2\pi/4$) when periodic impurities (triangles) with $a_\text{imp} = 8$ are present. The spin and the chirality structure factors are shown on the right:
(e) ferrichiral 3$Q^M$-2$Q^\chi$ spiral II ($H/J_3=2.2$ and $T/J_3=0.35$), (f) antiferrochiral single-$Q^M$ spiral ($H/J_3=1.0$ and $T/J_3=0.35$), (g) ferrochiral 3$Q^M$-2$Q^\chi$ spiral ($H/J_3=0.1$ and $T/J_3=0.05$), and (h) vertical single-$Q^M$ spiral ($H/J_3=0.2$ and $T/J_3=0.35$). 
 The circles with solid (dashed) lines indicate the dominant (subdominant) $\mathbf{q} \ne 0$ peak(s).
  Note that the $\mathbf{q}=0$ component is removed from $S_s^{zz}(\mathbf{q})$.
  The hexagon with a solid (dashed) line shows the first Brillouin zone (of the impurity superlattice).
  To obtain the smooth spin configurations, we integrate out short wavelength fluctuations by averaging over 50-500 MCS.
}
\end{figure*}

\subsection{Case with $a_\text{imp} = 8$}
\label{sec:a8}

The configuration of impurities is shown in the inset of Fig.~\ref{Fig:phase-diagram_a8}(a).
The impurities are separated by $a_\text{imp} = 8$ sites along the lattice principal directions.
This superlattice spacing is exactly twice as large as $2\pi/ Q =4$.

Figure \ref{Fig:phase-diagram_a8}(a) shows the $T$-$H$ phase diagram obtained with our MC simulations, which features eight different phases other than the paramagnetic state (Table~\ref{tab_a8}). The phase boundaries are determined by analyzing the structure factors in Eqs.~\eqref{Eq:spinstructure} and \eqref{Eq:chiralstructure} (Figs.~\ref{Fig:phase1-4_a8} and \ref{Fig:phase5-8_a8}) and the peak in the uniform magnetic susceptibility [Fig.~\ref{Fig:phase-diagram_a8}(d)] as a function of $H$.
Notably, many phases are multiple-$Q$ states that support long-wavelength modulation of local scalar chirality (chirality wave), which is not necessarily single-$Q^\chi$ but some of them are actually of the multiple-$Q^\chi$ type (our convention is to use $Q^{M}$ and $Q^{\chi}$, respectively, when it is necessary to make an unambiguous distinction between spin and chirality textures).
The commensuration  between the chirality waves and  the impurity superlattice leads to a majority of magnetically ordered phases with  net scalar spin chirality [see Figs.~\ref{Fig:phase-diagram_a8}(c) and \ref{Fig:phase-diagram_a8}(e)].
The uniform component arises from uncompensated positive and negative components of the chirality wave texture: the impurities remove spins contributing to only one sign of the modulated chiral structure. Meanwhile, the $xy$ spin components only exhibit   quasi-long-range correlations  in $d = 2$, as expected from Mermin-Wagner's theorem.~\cite{Mermin_PhysRevLett.17.1133}
In agreement with our discussion above, the spins around the impurities form vortices with  enhanced $xy$ components. 
In what follows, we describe details of the obtained phases.

\paragraph{Ferrochiral 3$Q^M$-6$Q^\chi$ vortex crystal~\footnote{In our convention, we count the number of pairs $\pm Q^M$ and $\pm Q^\chi$ corresponding to the (quasi-)Bragg peaks in the structure factor.}}
This phase appears right below the saturation field near $T = 0$. Upon entering this phase via the thermal transition, the specific heat shows a single weak anomaly as shown in Fig.~\ref{Fig:phase-diagram_a8}(f). The spin configuration is a triple-$Q^M$ vortex crystal similar to the one reported recently in 3D frustrated quantum magnets~\cite{Kamiya_PhysRevX.4.011023,Wang_PhysRevLett.115.107201}. 
This phase may be understood as the natural extension of the single-impurity vortex for $\hsat < H < \hsat^I$~\cite{Lin_PhysRevLett.116.187202}. 
In fact, as shown in Fig.~\ref{Fig:phase1-4_a8}(a), each nonmagnetic impurity induces a vortex very similar to the schematic picture in Fig.~\ref{Fig:vortex}, forming a triangular vortex lattice as a whole. Upon closer inspection, it is possible to see that antivortices appear right in the middle of nearest neighbor vortices, so that the ``boundary condition" by the impurity array is satisfied.
The vortices and antivortices give opposite contributions to the  scalar spin chirality.
While this would lead to a total cancellation of $\chi^\text{tot}$ in a system without impurities~\cite{Kamiya_PhysRevX.4.011023}, the present vortex crystal has net scalar spin chirality because the local scalar chirality around each impurity  has the same sign. 
The net chirality for up and down triangles is equal, $\chi^u = \chi^d$, which we refer to as ``ferrochiral.''
Interestingly, the chirality wave is primarily characterized by higher harmonics relative to the wave vectors that minimize $J(\mathbf{q})$, namely, $\mathbf{Q}_1^\chi = \mathbf{Q}_1 + 2\mathbf{Q}_2$, etc.

\paragraph{Ferrochiral 3$Q^M$ vortex crystal}
This phase can be seen as another type of crystallization of vortices found in Ref.~\onlinecite{Lin_PhysRevLett.116.187202} for $\hsat < H < \hsat^I$. It occupies a small corner of the entire region of the ordered phases, which appears right below the saturation field in the range of intermediate temperature, $0.37 \lesssim T / J_3 \lesssim 0.43$.
The spin configuration is characterized by the triple-$Q^M$ modulation and net spin scalar chirality ($\chi^u = \chi^d \ne 0$) [Fig.~\ref{Fig:phase1-4_a8}(b)].
However, unlike the ferrochiral 3$Q^M$-6$Q^\chi$ vortex crystal phase discussed above, the chirality texture does not support a static finite-$Q^\chi$ component in the thermodynamic limit (namely, the chirality texture is homogeneous) as shown in Fig.~\ref{Fig:sizedep_opt1-4}(b) in Appendix~\ref{sec:Finite-size scaling of each phase}; 
this aspect distinguishes  this phase from the skyrmion crystal phase~\cite{Okubo_PhysRevLett.108.017206}, although both phases are triple-$Q^M$ and chiral. Moreover, the $z$ component near the impurities is $S^z \sim 0$ in the ferrochiral 3$Q^M$ vortex crystal phase, while it is $S^z \sim -1$ in the skyrmion crystal phase, which implies a different topological nature. 
The scalar chirality shows a peak near the phase boundary between this and the ferrichiral 3$Q^M$-2$Q^\chi$ spiral II phase discussed below [Fig.~\ref{Fig:phase-diagram_a8}(c)]. 

\paragraph{Ferrochiral 3$Q^M$-1$Q^\chi$ spiral}
This phase occupies a smaller $H$ region next to the ferrochiral 3$Q^M$-6$Q^\chi$ vortex crystal phase at low $T$. As shown in Fig.~\ref{Fig:phase1-4_a8}(c), it has a triple-$Q^M$ spin texture.  
In the structure factor for the $xy$ component, there are two dominant peaks and a single subdominant peak, while a single-$Q$ peak is in the structure factor for the $z$ component. 
Upon closer looking, it becomes clear that the spin texture is pinned by the impurities via
the parts with local chirality having 
the same sign; for this reason this state is chiral, $\chi^u = \chi^d \ne 0$.
In addition to the uniform component, the chirality texture has a single-$Q^\chi$ component corresponding to the stripe modulation. 

\paragraph{Ferrichiral 3$Q^M$-2$Q^\chi$ spiral I}
This phase occupies the largest portion of the phase diagram among the ordered phases and it is found next to the ferrochiral 3$Q^M$-1$Q^\chi$ spiral upon decreasing $H$. 
The spin configuration is characterized by the triple-$Q^M$ noncoplanar modulation. $S^{zz}_s(\mathbf{q})$ has two dominant peaks with an additional smaller peak, which in $S^{\perp}_s(\mathbf{q})$ in turn correspond to two subdominant peaks and the major peak, respectively [Fig.~\ref{Fig:phase1-4_a8}(d)].
We find that the impurities are on the contour $S^z \approx 0$ as shown in the left panel of Fig.~\ref{Fig:phase1-4_a8}(d) in accordance with the general argument that the effect of the magnetic field is effectively weakened for spins surrounding impurities.
The chirality wave is mainly characterized by the double-$Q^\chi$ modulation, which can be seen as the ``checkerboard" pattern. Note that the ferrichiral 3$Q^M$-2$Q^\chi$ spiral I phase also possesses a small subdominant peak at $\mathbf{Q}_1+\mathbf{Q}_2$ as shown in the chiral structure factor in Fig.~\ref{Fig:phase1-4_a8}(d). 
A subtle difference is that $S^{u}_{\chi}(\mathbf{q})$ and $S^{d}_{\chi}(\mathbf{q})$ have different profiles in this state. Based on this observation, we call this state ``ferrichiral.''

\paragraph{Ferrichiral 3$Q^M$-2$Q^\chi$ spiral II}
This state is found in the intermediate-$H$ regime, next to the ferrichiral 3$Q^M$-2$Q^\chi$ spiral I phase upon increasing $T$. While this is very similar to the ferrichiral 3$Q^M$-2$Q^\chi$ spiral I, the intensities of the two dominant components of the chiral structure factor are different in this phase [Fig.~\ref{Fig:phase5-8_a8}(e)] , while the ones for  the ferrichiral 3$Q^M$-2$Q^\chi$ spiral I phase are the same. 

\paragraph{Antiferrochiral single-$Q^M$ spiral}
This state appears next to the ferrichiral 3$Q^M$-2$Q^\chi$ spiral II state upon decreasing $H$. The spin configuration shown in Fig.~\ref{Fig:phase5-8_a8}(f) resembles the conical spiral state
that is obtained without impurities [Fig.~\ref{Fig:phase-diagram_a8}(b)]. The difference, however, is that the impurities introduce the additional weak longitudinal modulation. The $C_6$ symmetry is broken in this phase as in the single-$Q$ spiral phase in the pure $J_1$-$J_3$ model with easy-plane anisotropy, where the symmetry of the global spin rotation,  U(1), is the same as in the present case.~\cite{tamura2011first} The obtained specific heat curve [Fig.~\ref{Fig:phase-diagram_a8}(g)] is consistent with the single first order phase transition reported by Tamura \etal~\cite{tamura2011first} in the pure easy-plane model, although a more careful finite size scaling is required to settle this point.
This state is not chiral because $\chi^u$ and $\chi^d$ cancel each other out: $\chi^u = -\chi^d$. For this reason we refer to this phase as ``antiferrochiral.''

\paragraph{Ferrochiral 3$Q^M$-2$Q^\chi$ spiral}
This phase occupies the low-field and low-$T$ region of the phase diagram [Fig.~\ref{Fig:phase-diagram_a8}(a)] and appears next to the ferrichiral 3$Q^M$-2$Q^\chi$ spiral I phase with decreasing $H$. The spin configuration [see Fig.~\ref{Fig:phase5-8_a8}(g)] closely resembles a single-$Q^M$ vertical spiral state (see below) though the small additional $\mathbf{Q}^M$ components render the spin configuration noncoplanar with the triple-$Q^M$ modulation. Meanwhile, the chirality wave texture shows the double-$Q^\chi$ modulation with a single-$Q^\chi$ subdominant component, which is very similar to that of the ferrichiral 3$Q^M$-2$Q^\chi$ spiral I state. 
The difference is that the net chirality for up and down triangles in the present state is equal, $\chi^u = \chi^d$, while it is different in the ferrichiral 3$Q^M$-2$Q^\chi$ spiral state.
Once again, it is evident that the spin texture is pinned by the impurities where the local chirality has the same sign, which leads to nonzero uniform scalar chirality.
The ferrochiral 3$Q^M$-2$Q^\chi$ spiral state extends its stability down to $H = 0$, as shown in Fig.~\ref{Fig:phase-diagram_a8}(e). 

\paragraph{Vertical single-$Q^M$ spiral}
This phase occupies a region at higher $T$ next to the ferrochiral 3$Q^M$-2$Q^\chi$ spiral phase.
This is a single-$Q^M$ coplanar state in an arbitrary plane containing the vertical  $c$-axis  when the magnetic field is applied in the $c$ direction. Thus, this state has no net chirality. The difference relative to the ferrochiral 3$Q^M$-2$Q^\chi$ spiral state  is the disappearance of the subdominant components in the spin structure factor induced by thermal fluctuations [Fig.~\ref{Fig:phase5-8_a8}(h)]. The transition from the paramagnetic state is suggested to be a single-step first-order phase transition (not shown). This is similar to the case in the pure $J_1$-$J_3$ model,~\cite{Tamura2008,tamura2011first,tamura2014phase} albeit with a subtle difference that the state in the latter is the conical spiral. 

\subsection{Case with $a_\text{imp} = 4$ 
  \label{sec:a4}
}
\setcounter{paragraph}{0}

\begin{table*}
\begin{center}
  \caption{
    Classification of the ordered phases for $J_3 / \lvert{J_1}\rvert \approx 0.854$ in the model with periodic impurities ($a_\text{imp} = 4$).
  }
  \vspace{2pt}
   \scalebox{0.85}{
  \begin{tabular}{lccccc}
    \toprule
    \hline \hline
    phase & \parbox[t]{65pt}{nonzero net \\scalar chirality} &
    \parbox[t]{90pt}{number of (quasi-)Bragg peaks in $S_s^{\perp}(\mathbf{Q})$} &
    \parbox[t]{90pt}{number of (quasi-)Bragg peaks in $S_s^{zz}(\mathbf{Q})$\\ at $\mathbf{q} \ne 0$}  & \parbox[t]{90pt}{number of Bragg peaks in $S_\chi^{u/d}(\mathbf{Q})$ at $\mathbf{q} \ne 0$}&
    \parbox[t]{90pt}{broken point-group symmetry (of the lattice with impurities)}\\
    \midrule
    \hline
    Nonchiral 2$Q^M$-3$Q^\chi$ vortex crystal & No & 2 & 0 & 2+1 & $C_6$ \\
    Nonchiral 3$Q^M$-2$Q^\chi$ spiral & No & 2 & 1 & 1+1 & $C_6$ \\
    \hline\hline
    \bottomrule
  \end{tabular}
  }
\label{tab_a4}
\end{center}
\end{table*}

\begin{figure}
\begin{center}
  \includegraphics[width=1.0\hsize,bb=0 0 788 530]{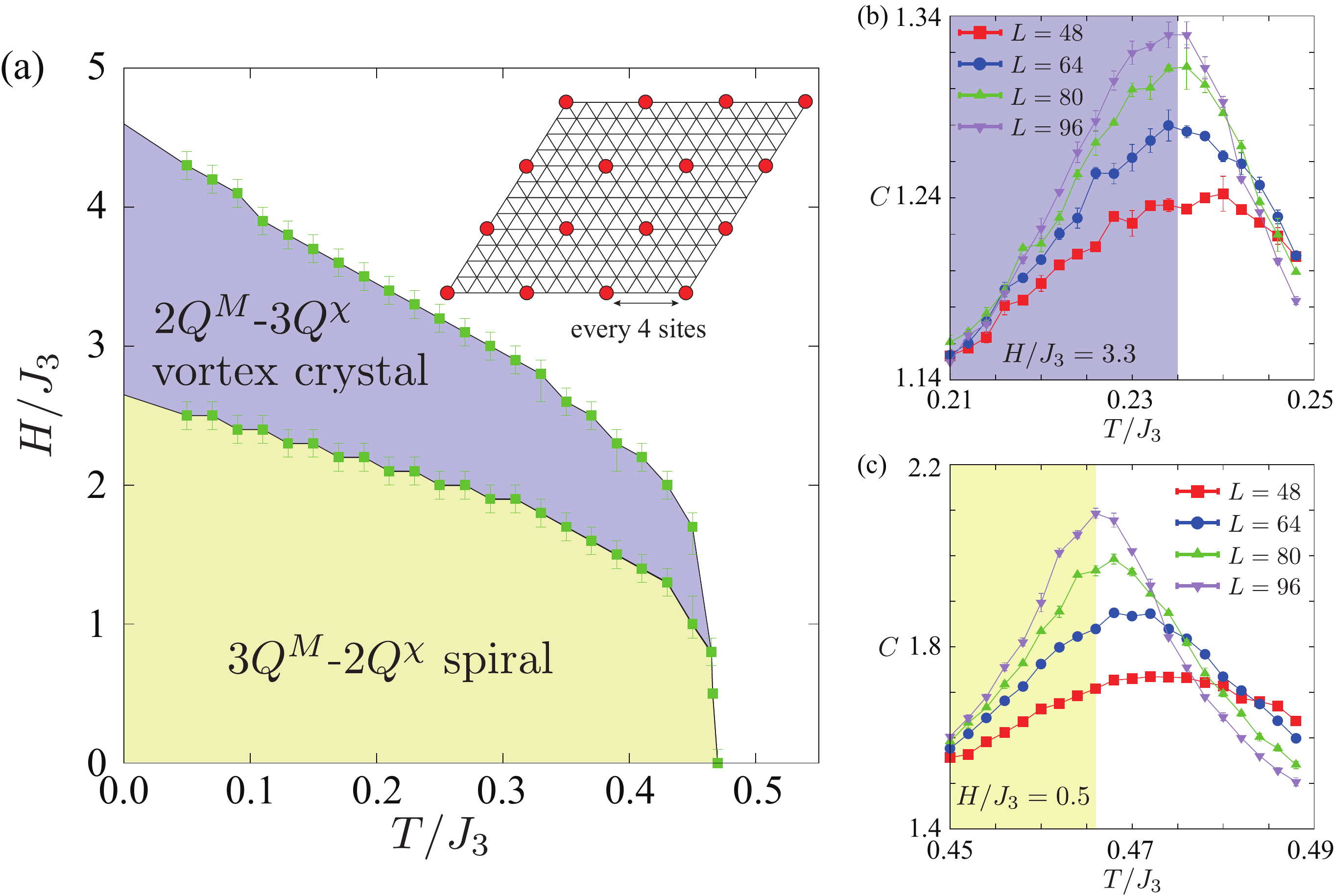} 
  \caption{
    \label{Fig:phase-diagram_a4}
    (a) Phase diagram of the model with a  periodic array of impurities  with $a_\text{imp} = 4$ and $J_3 / |J_1| \approx 0.854$ ($Q = 2\pi/4$). The inset shows the impurity configuration.    
    (b), (c) $T$ dependence of the specific heat at (b) $H/J_3=3.3$ and (c) $H/J_3=0.5$.
   }
\end{center}
\end{figure}

\begin{figure*}
\begin{center}
  \includegraphics[width=0.85\hsize, bb=0 0 740 535]{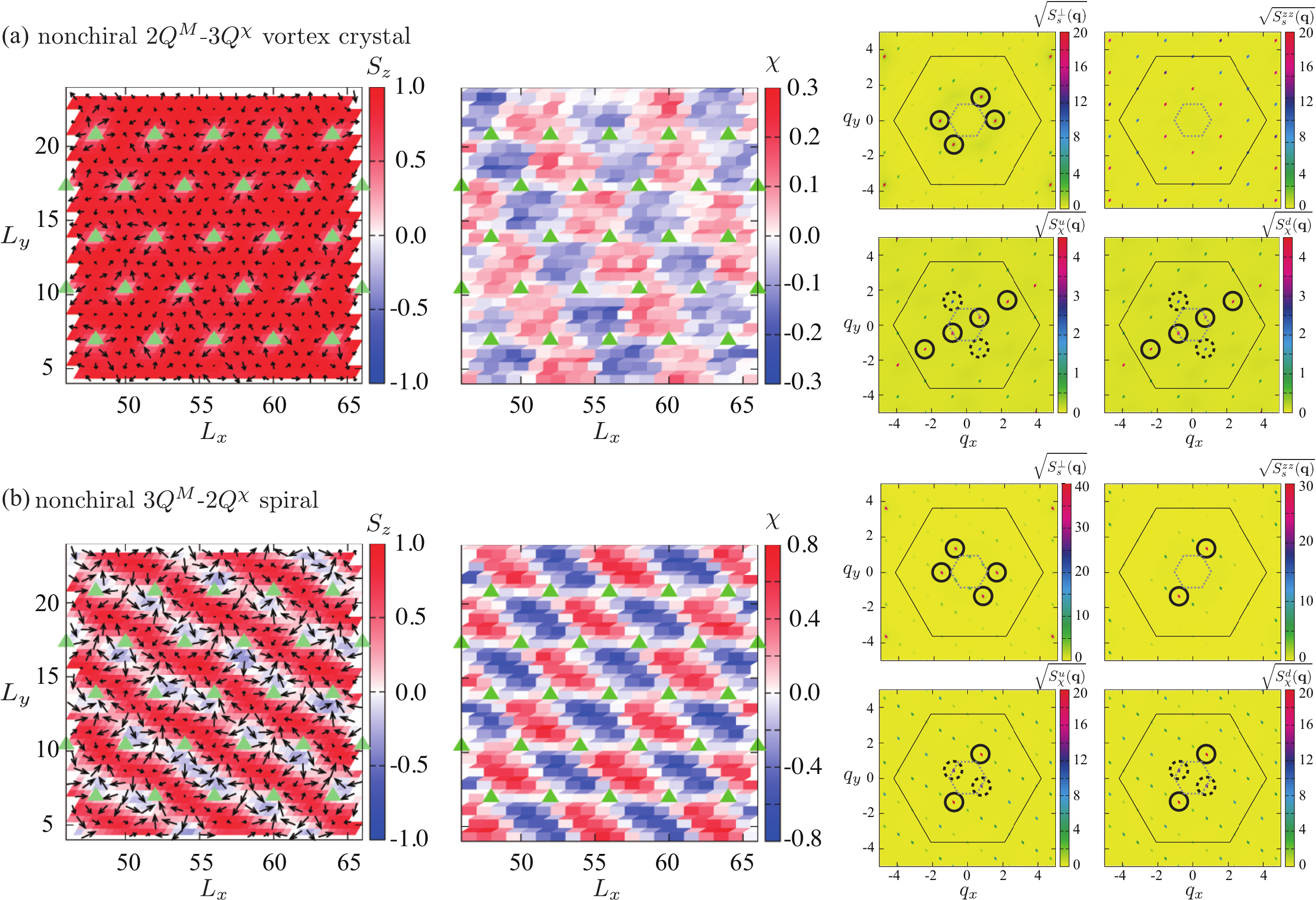}
\end{center}  
\caption{
  \label{Fig:phases_a4}
  Snapshots of the spin configuration (left), the chirality texture (middle), and the spin and the chirality structure factors (right) in the model with periodic impurities (triangles) with $a_\text{imp} = 4$ for $J_3 / \lvert{J_1}\rvert \approx 0.854$ ($Q = 2\pi/4$): (a) nonchiral 2$Q^M$-3$Q^\chi$ vortex crystal  ($H/J_3=3.9$) and (b) nonchiral 3$Q^M$-2$Q^\chi$ spiral  ($H/J_3=1.6$). The circles with solid (dashed) lines indicate the dominant (subdominant) peak(s) at $\mathbf{q} \ne 0$.
  The $\mathbf{q}=0$ component is removed from $S_s^{zz}(\mathbf{q})$.
  The hexagon with a solid (dashed) line shows the first Brillouin zone (of the impurity superlattice).
We average over 50-500 MCS to integrate out short wavelength fluctuations.
}
\end{figure*}

Next, we briefly discuss the case where the superlattice constant $a_\text{imp} = 4$ is half of the previous case, as illustrated in the inset of Fig.~\ref{Fig:phase-diagram_a4}(a). This corresponds to the relation $\mathbf{K}_+ + \mathbf{K}_- = 2\mathbf{Q}_1$. Even though the  commensurability still holds, the obtained phases summarized in Table~\ref{tab_a4} are very different from the previous case, except for the fact that a vortex crystal still appears right below the saturation field.

\paragraph{Nonchiral 2$Q^M$-3$Q^\chi$ vortex crystal}
As it is shown in Fig.~\ref{Fig:phases_a4}(a), the high field state is  another vortex crystal, in which both the vortices and anti-vortices are nucleated by the impurities. The vortex and antivortex-chains alternate creating the stripe pattern shown in the middle of Fig.~\ref{Fig:phases_a4}(a). This stripe pattern breaks the  translational symmetry of the system.
Vortices and antivortices have opposite scalar spin chirality, producing a chirality wave with a
wave vector equal to half of the impurity lattice reciprocal vector: $\mathbf{Q}^\chi = \mathbf{K}^+ / 2$ [Fig.~\ref{Fig:phases_a4}(a)]. Given that both vortex structures, the vortices and the antivortices,  are nucleated around the impurities, the net chirality is perfectly cancelled, rendering this state non-chiral. Figure~\ref{Fig:phase-diagram_a4}(b) shows the specific curve heat near the thermal phase transition. Though it is suggestive of a single-step continuous phase transition, a more careful analysis will be required to draw a final conclusion about the critical behavior.

\paragraph{Nonchiral 3$Q^M$-2$Q^\chi$ spiral}
The spin configuration in the low field phase has a single-$Q^M$ longitudinal modulation and a double-$Q^M$ transverse modulation [Fig.~\ref{Fig:phases_a4}(b)].
A closer inspection reveals that both $S^{\perp}_s(\mathbf{q})$ and $S^{zz}_s(\mathbf{q})$ have additional peaks induced by the impurities, such as the peaks at $q=\pi/(\sqrt{3}) \approx 1.814$ and $q=\pi/(2\sqrt{3}) \approx 0.907$. This fact is more evident for the double-$Q^\chi$ chirality wave texture, in which the subdominant component is induced by the impurities.
The spin configuration accommodates itself to the impurity superlattice in such a way that the impurities are on the nodal line of the chirality wave texture [Fig.~\ref{Fig:phases_a4}(b)]. As depicted in Fig.~\ref{Fig:phases_a4}(b), the net chirality vanishes. 
The specific heat near the thermal phase transition is suggestive of a single-step continuous phase transition also in this case [Fig.~\ref{Fig:phase-diagram_a4}(c)]. 

\section{Variational analysis
  \label{sec:variational}
}
Our numerical calculation indicate that multiple-$\mathbf{Q}$ ground states are realized instead of the single-${\boldmath Q}$ conical ground state that is obtained for the pure system.
Below we provide a variational analysis confirming that the commensurability relation between the ${\boldmath Q}$-vectors and the superlattice reciprocal vectors $\mathbf{K}_\pm$ renders the single-$Q$ conical  state unstable towards  more complex multiple-$\mathbf{Q}$ structures.
For the sake of concreteness, we consider below the case $a^{}_\text{imp} = 8$ where $\mathbf{K}^{}_{+} + \mathbf{K}^{}_{-} = \mathbf{Q}_1$.

First we note that $\ham^{J}_\text{imp}$ in  Eq.~\eqref{eq:Himp} can be written as
\begin{align}
  \ham^{J}_\text{imp} = -\rho^{}_\text{imp} \sum_{\mathbf{q},\mathbf{q}' \in \text{FBZ}} 2J(\mathbf{q})
  \left(\sum_{\mathbf{G}_\text{imp}} \delta_{\mathbf{q}', \mathbf{q} + \mathbf{G}_\text{imp}}\right)
  \mathbf{S}_{\mathbf{q}} \cdot \mathbf{S}_{-\mathbf{q}'},
  \label{eq:Himp:2}
\end{align}
after taking the Fourier transform $\mathbf{S}_\mathbf{q} = \sqrt{{1}/{N}} \sum_j e^{-i\mathbf{q}\cdot\mathbf{r}_j} \mathbf{S}_j$.
$\mathbf{G}_\text{imp}$ runs over the set of impurity superlattice reciprocal vectors and $\rho^{}_\text{imp} = a^{-2}_\text{imp}$ is impurity concentration. Thus, because of the commensurability relation, $\ham^{J}_\text{imp}$ couples the different ${\mathbf Q}$ vectors and the $\mathbf{q} = 0$ component  induced by the magnetic field. Likewise, $\ham^{H}_\text{imp}$ is written as
\begin{align}
  \ham^{H}_\text{imp} =  \sqrt{N} \rho^{}_\text{imp} H \sum_{\mathbf{G}_\text{imp} \in \text{FBZ}} S_{\mathbf{G}_\text{imp}}^z,
\end{align}
where the $Q$ vectors and $\mathbf{q} = 0$ are both included in the summation.

\subsection{Stability analysis of the conical spiral}
We start from showing that the impurity-induced coupling makes the single-${\boldmath Q}$ state unstable at $T = 0$. To this end, we consider  the following deformation of the single-${\boldmath Q}$ conical state, which   satisfies the fixed-length constraint required for classical spins:
\begin{align}
  S_j^x &= \sqrt{\sin^2\tilde{\theta} - \delta^2} \cos(\mathbf{Q}_1\cdot\mathbf{r}_j) + \delta \cos(\mathbf{Q}_2\cdot\mathbf{r}_j),
  \notag\\
  S_j^y &= \sqrt{\sin^2\tilde{\theta} - \delta^2} \sin(\mathbf{Q}_1\cdot\mathbf{r}_j) - \delta \sin(\mathbf{Q}_2\cdot\mathbf{r}_j),
  \notag\\
  S_j^z &= \sqrt{\cos^2\tilde{\theta} - 2\delta \sqrt{\sin^2\tilde{\theta} - \delta^2} \cos(\mathbf{Q}_3\cdot\mathbf{r}_j)},
  \label{eq:ansatz}
\end{align}
where $\delta$ parametrizes the magnitude of the deformation  and $\tilde{\theta}$ [which is equal to $\cos^{-1}(S_j^z),~\forall j$ for $\delta \to 0$] is a variational parameter.
Following Ref.~\onlinecite{Hayami_PhysRevB.93.184413}, we introduce
\begin{align}
  x = \delta \cos^{-2}\tilde{\theta} \sqrt{\sin^2\tilde{\theta} - \delta^2},
\end{align}
and $S_j^z$ can be expanded as
\begin{align}
  S_j^z = \cos\tilde{\theta} \sum_{n\ge 0} f_n(x) \cos\left(n\mathbf{Q}_3 \cdot \mathbf{r}_j\right),
\end{align}
where, to $\bigO{\delta^5}$, $f_0(x) = 1 - \textfrac{x^2}{4} - \textfrac{15x^4}{64}$, $f_1(x) = -x - \textfrac{3x^3}{8}$, $f_2(x) = -\textfrac{x^2}{4} - \textfrac{5x^3}{16}$, $f_3(x) = - \textfrac{x^3}{8}$, $f_4(x) = - \textfrac{5x^4}{64}$, and $f_{n \ge 5}(x)$ can be neglected at this order.
We find
\begin{align}
  \left\langle{\ham^{H}_\text{imp}}\right\rangle
  &=  N \rho^{}_\text{imp} H \cos \tilde{\theta} \left(1 - x - \frac{x^2}{2} - \frac{x^3}{2} - \frac{35 x^4}{64} \right)
  \notag\\
  &+ \bigO{\delta^5}.
\end{align}
Also, by splitting Eq.~\eqref{eq:Himp:2} into different spin components as $\ham^{J}_{\text{imp}} = \ham_{\text{imp}}^{xx} + \ham_{\text{imp}}^{yy} + \ham_{\,\text{imp}}^{zz}$ and denoting $J_{n Q} = J(n\mathbf{Q}_1) = J(n\mathbf{Q}_2) = J(n\mathbf{Q}_3)$ and $J_0 = J(0)$,
we find
\begin{align}
  \left\langle{\ham_{\text{imp}}^{xx}}\right\rangle
  &= -2 N \rho^{}_\text{imp} J_Q \left( \sin^2\tilde{\theta} + 2 x \cos^2\tilde{\theta} \right),
  \notag\\
    \left\langle{\ham_{\text{imp}}^{yy}}\right\rangle
    &= 0,
\end{align}
which are independent of the deformation, and
\begin{widetext}
\begin{align}
    \left\langle{\ham_{\text{imp}}^{zz}}\right\rangle
    &= -2 N \rho^{}_\text{imp} \cos^2\tilde{\theta}
    \left[
      \frac{1}{4} J_0 f^2_0(x)
      + \frac{1}{2} f_0(x) \sum_{n \ge 1} \left( J_0 + J_{n Q} \right) f_n(x)
      + \sum_{n \ge 1} J_{nQ} f_n(x) \sum_{m \ge 1} f_m(x)
      \right]
    \notag\\
    &= -2 N \rho^{}_\text{imp} \cos^2\tilde{\theta}~
    \Biggl[
      \frac{J_0}{4}
      - \frac{J_0 + J_{Q}}{2} x
      - \frac{2 J_0 - 8 J_{Q} + J_{2Q}}{8} x^2
      + \frac{-2 J_0 + 3 J_{Q} + 4 J_{2Q} - J_{3Q}}{16} x^3
      \notag\\[-2pt]
      &\hspace{185pt}
      + \frac{-34 J_0 + 112 J_Q - 8 J_{2Q} + 16 J_{3Q} - 5 J_{4Q}}{128} x^4
      \,\Biggr]
    + \bigO{\delta^5}.
  \end{align}
\end{widetext}
These results show that $\left\langle{\ham_{\,\text{imp}}}\right\rangle_\delta - \left\langle{\ham_{\,\text{imp}}}\right\rangle_{\delta=0}$ includes a  \textit{linear} contribution  in the deformation parameter $\delta$. In the meantime, the change in $\ham_\text{pure}$ was evaluated in Ref.~\onlinecite{Hayami_PhysRevB.93.184413} as $\left\langle{\ham_{\text{pure}}}\right\rangle_\delta - \left\langle{\ham_{\text{pure}}}\right\rangle_{\delta=0} = N \cos^2\tilde{\theta}\,(J_{2Q} - J_0) x^4 /32 + \bigO{\delta^5}$. Thus,  $\left\langle{\ham}\right\rangle_\delta - \left\langle{\ham}\right\rangle_{\delta=0}$ decreases linearly in $\delta$, implying that the conical spiral is indeed unstable in the presence of  periodic array of impurities, which is commensurate with the ordering wave vectors.

\subsection{Luttinger-Tisza analysis}
The next question is whether the numerically found field-induced phases can be analytically explained in a simple manner.
Below we first perform a soft-spin variational analysis at $T = 0$ by adopting the following ansatz,
\begin{align}
  \tildeS_j
  &= \mathbf{M}_{0} + \sum_{1 \le \mu \le 3} \left( \mathbf{M}_{\Qmu} e^{i\mathbf{Q}_\mu \cdot \mathbf{r}_j} + \text{c.c.}\right),
\end{align}
where $\mathbf{M}_{0} = N^{-1} \sum_{j} \tildeS_j$ is a three-component real vector for the uniform component and $\mathbf{M}_{\Qmu} = N^{-1}\sum_{j} e^{-i\mathbf{Q}_{\mu}\cdot\mathbf{r}_j}\tildeS_j$ ($1\le\mu\le3$) are three-component complex vectors. The tilde attached to the spin variable indicates that the fixed-length constraint is replaced by the average normalization condition given by a quadratic function
\begin{align}
  N^{-1} \sum_j \left\lvert{\tildeS_j}\right\rvert^2
  &= \vsq{\mathbf{M}_{0}} + 2 \sum_{1 \le \mu \le 3} \vsq{\mathbf{M}_{\Qmu}} = 1.
\end{align}
This average constraint can be easily taken into account with the Lagrange multiplier method.

The variational energy density extended by the Lagrange multiplier $\lambda$ is $E_\text{var} = E_J^\text{pure} + E_J^\text{imp} + E_{H} + E_{\lambda}$ with
\begin{align}
  E_J^\text{pure}
  &= J_0 \vsq{\mathbf{M}_{0}} + 2 J_Q \sum_{1 \le \mu \le 3} \vsq{\mathbf{M}_{\Qmu}},
  \notag\\
  E_J^\text{imp}
  &=
  -2 \rho^{}_\text{imp} J_0 \vsq{\mathbf{M}_{0}}
  \notag\\
  &\hspace{0pt} -2 \rho^{}_\text{imp} J_Q \vsq{\sum_{1 \le \mu \le 3} \left( \mathbf{M}_{\Qmu} + \mathbf{M}^{\ast}_{\Qmu} \right)}
  \notag\\
  &\hspace{0pt} -2 \rho^{}_\text{imp} \left(J_0 + J_Q\right)  \sum_{1 \le \mu \le 3}
  \mathbf{M}_{0} \cdot
  \left( \mathbf{M}_{\Qmu} + \mathbf{M}^{\ast}_{\Qmu} \right),
  \allowdisplaybreaks[2]
  \notag\\
  E_H
  &=- (1 - \rho^{}_\text{imp}) H M^{z}_{0}
  \notag\\
  &\hspace{10pt}+ \rho^{}_\text{imp} H \sum_{1 \le \mu \le 3} \left[ M^z_{\Qmu} + \bigl(M^z_{\Qmu}\bigr)^\ast \right],
  \allowdisplaybreaks[2]
  \notag\\
  E_{\lambda}
  &= - \lambda \left(\vsq{\mathbf{M}_{0}} + 2 \sum_{1 \le \mu \le 3} \vsq{\mathbf{M}_{\Qmu}} - 1 \right).
\end{align}
By rewriting $\mathbf{M}_0 = \mathbf{A}_0$ and $(\Re{\mathbf{M}_{\Qmu}},\Im{\mathbf{M}_{\Qmu}}) = (\mathbf{A}_{\mu}, \mathbf{B}_{\mu})$ for $1 \le \mu \le 3$, we first look into the quadratic part $E_\text{var}^\text{quad} = E_\text{var}(\lambda) - E_H$,
\begin{align}
  E_\text{var}^\text{quad}
  &= 
  \begin{pmatrix}
    \mathbf{A}_0 &
    \mathbf{A}_1 &
    \mathbf{A}_2 &
    \mathbf{A}_3
  \end{pmatrix}
  \begin{pmatrix}
    \omega^{}_{0} & \Delta_{0Q} & \Delta_{0Q} & \Delta_{0Q}
    \\
    \Delta_{0Q} & \omega^{}_Q & \Delta_Q & \Delta_Q
    \\
    \Delta_{0Q} & \Delta_Q & \omega^{}_Q & \Delta_Q
    \\
    \Delta_{0Q} & \Delta_Q & \Delta_Q & \omega^{}_Q
  \end{pmatrix}
  \begin{pmatrix}
    \mathbf{A}_0 \\
    \mathbf{A}_1 \\
    \mathbf{A}_2 \\
    \mathbf{A}_3
  \end{pmatrix}
  \notag\\
  &\hspace{10pt}+ 2 J_Q \sum_{1\le\mu\le3} \vsq{\mathbf{B}_\mu}
  \notag\\
  &= \sum_{0 \le \kappa \le 3}\eps_\kappa \vsq{ \bm{\Phi}_\kappa }
  + 2 J_Q \sum_{1\le\mu\le3} \vsq{\mathbf{B}_\mu},
\end{align}
where $\omega^{}_{0} = (1 - 2\rho^{}_\text{imp})J_0 - \lambda$, $\omega^{}_Q = (2 - 8\rho^{}_\text{imp})J_Q - 2 \lambda$, $\Delta_{0Q} = -2 \rho^{}_\text{imp} (J_0 + J_Q)$, and $\Delta_Q = -8\rho^{}_\text{imp} J_Q$.
Here we have diagonalized the $4\times4$ real symmetric coefficient matrix, obtaining the eigenvalues, $\eps_0 = \eps_1 = 2 (J_Q - \lambda)$, $\eps_2 = \eps^{}_+$, and $\eps_3 = \eps^{}_-$ with
\begin{align}
  \eps^{}_\pm = \frac{\omega^{}_0 + \omega^{}_Q + 2\Delta_Q \pm \sqrt{12 \Delta_{0Q}^2 + \left( \omega^{}_0 - \omega^{}_Q - 2\Delta_Q \right)^2}}{2}.
\end{align}
$\{\mathbf{A}_\mu\} \to \{\mathbf{\Phi}_{\kappa}\}$ is the associated orthogonal transformation,
\begin{align}
  \begin{pmatrix}
    \mathbf{\Phi}_0 \\[7pt]
    \mathbf{\Phi}_1 \\[7pt]
    \mathbf{\Phi}_2 \\[7pt]
    \mathbf{\Phi}_3
  \end{pmatrix}
  =
  \begin{pmatrix}
    0 & \frac{2}{\sqrt{6}} & -\frac{1}{\sqrt{6}} & -\frac{1}{\sqrt{6}}
    \\[7pt]
    0 &                  0 &  \frac{1}{\sqrt{2}} & -\frac{1}{\sqrt{2}}
    \\[7pt]
    \frac{c^{}_+}{\sqrt{c_+^2 + 3}} & \frac{1}{\sqrt{c_+^2 + 3}} & \frac{1}{\sqrt{c_+^2 + 3}} & \frac{1}{\sqrt{c_+^2 + 3}}
    \\[7pt]
    \frac{c^{}_-}{\sqrt{c_-^2 + 3}} & \frac{1}{\sqrt{c_-^2 + 3}} & \frac{1}{\sqrt{c_-^2 + 3}} & \frac{1}{\sqrt{c_-^2 + 3}}
  \end{pmatrix}
  \begin{pmatrix}
    \mathbf{A}_0 \\[7pt]
    \mathbf{A}_1 \\[7pt]
    \mathbf{A}_2 \\[7pt]
    \mathbf{A}_3
  \end{pmatrix},
  \label{eq:orth}
\end{align}
with
\begin{align}
  c^{}_{\pm}
  =  \frac{\eps^{}_\pm - \omega^{}_Q - 2\Delta_Q}{\Delta_{0Q}}.
\end{align}
By adding the Zeeman contribution, we have
\begin{align}
  E_\text{var}
  &= \sum_{0 \le \kappa \le 3}\eps_\kappa \norm{ \mathbf{\Phi}_\kappa }
  + 2 (J_Q - \lambda) \sum_{1\le\mu\le3} \norm{\mathbf{B}_\mu}
  \notag\\[-2pt]
  &\hspace{10pt}
  - \frac{(1 - \rho^{}_\text{imp})c^{}_+ - 6\rho^{}_\text{imp}}{\sqrt{c_+^2 + 3}} H {\Phi}^z_2 
  \notag\\
  &\hspace{10pt}
  - \frac{(1 - \rho^{}_\text{imp})c^{}_- - 6\rho^{}_\text{imp}}{\sqrt{c_-^2 + 3}} H {\Phi}^z_3.
\end{align}
Thus, by taking derivatives of $E_\text{var}$ with respect to $\Phi_\kappa^{x,y,z}$ and $B_\mu^{x,y,z}$, we can see that the magnetic field coupling to the triple-$Q$ modes ${\Phi}^z_{2,3}$ [see Eq.~\eqref{eq:orth}] leads to a  field-induced triple-$Q$ state within 
the Luttinger-Tisza approximation. However, the resulting  triple-$Q$ state is collinear, implying a rather strong violation of the fixed-length constraint.

\subsection{Real-space variational analysis}
\begin{figure}[t!]
  \begin{center}
    \includegraphics[width=0.9 \hsize,bb=0 0 545 526]{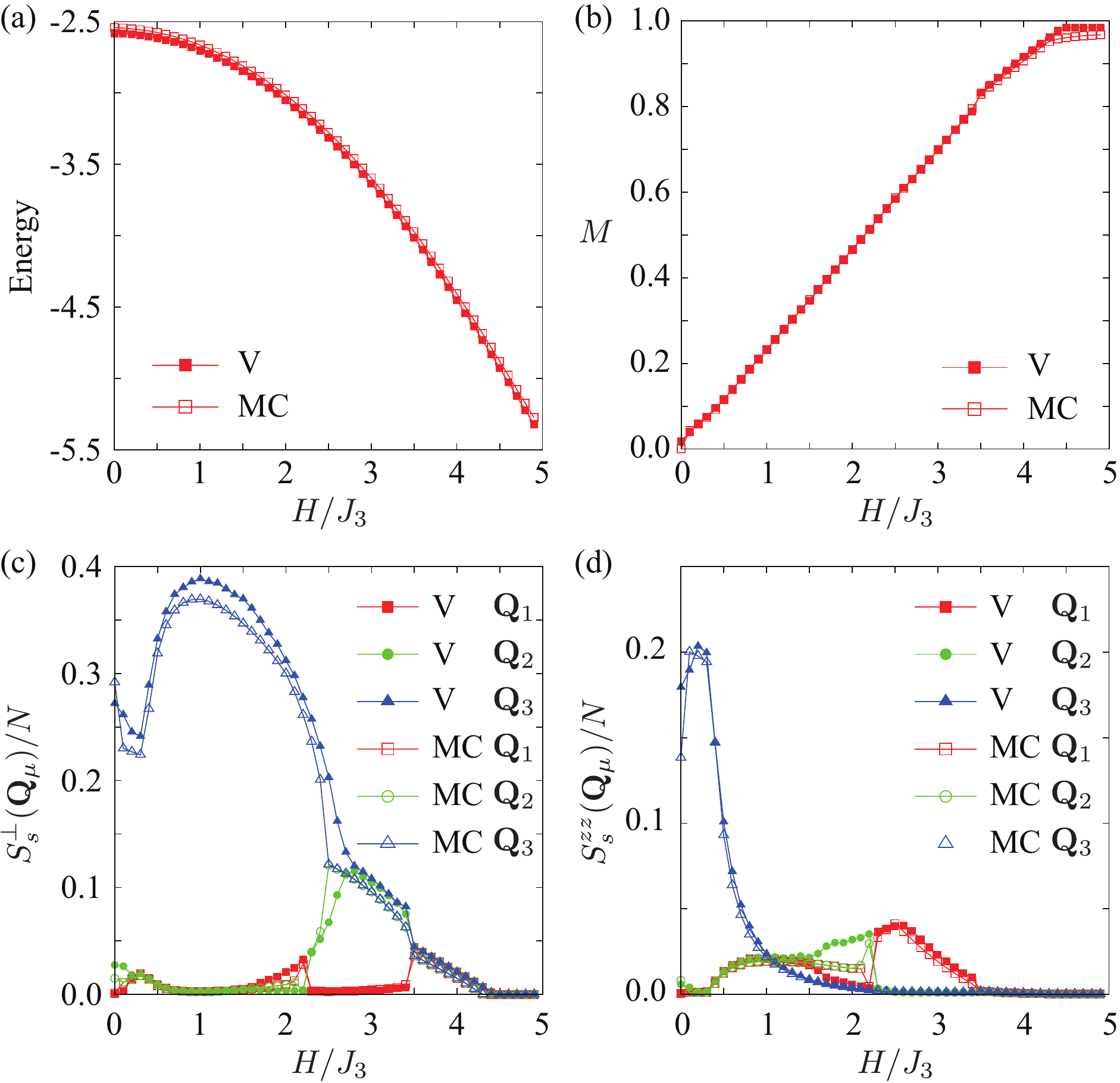} 
    \caption{
      \label{Fig:variational}
      $H$-dependences of (a) the energy per site, (b) magnetization, $M$, (c) $S^{\perp}_{s} (\mathbf{Q}_\nu)/N$, and (d) $S^{zz}_{s} (\mathbf{Q}_\nu)/N$ ($\nu=1-3$)  for $Q=2\pi/4$ and $a_{\rm imp}=8$. ``V" and ``MC" represent the results obtained by $T=0$ variational calculations and Monte Carlo simulations at $T/J_3=0.05$. 
    }
  \end{center}
\end{figure}

The above Luttinger-Tisza analysis suggests that a more strict treatment of the constraint is crucial, and the corresponding nonlinear effect is expected to drive the collinear triple-${\mathbf Q}$ state into a noncoplanar multi-${\mathbf Q}$ state. To proceed, we note that the commensurate ${\mathbf Q}$ vectors allow to work on a real-space variational calculation based on the following simple ansatz,
\begin{equation}
  {\bm S}_j = \frac{1}{N_j}
  \left[
    \mathbf{M}_{0} + \sum_{1 \le \mu \le 3} \left( \mathbf{M}_{\Qmu} e^{i\mathbf{Q}_\mu \cdot \mathbf{r}_j} + \text{c.c.}\right)
    \right],
\label{var}
\end{equation}
where $N_j$ is a normalization factor which enforces the ${\bm S} \cdot {\bm S} =1$ constraint exactly and $\mathbf{M}_{\mu}=\mathbf{M}^*_{-\mu}$. 
The variational parameters are seven: $M_0$ and three pairs of (Re$\mathbf{M}_\mu$, Im$\mathbf{M}_\mu$) ($\mu=1$-$3$).  

Figure~\ref{Fig:variational} shows the $H$ dependence of (a) the energy density, (b) magnetization, (c) the $xy$ component of the spin structure factor, and (d) the $z$ component of the spin structure factor, which are obtained by variational calculations at $T/J_3=0.00$ and Monte Carlo simulations at $T/J_3=0.05$. 
The  Monte Carlo calculations are consistent with the variational results except for the region $1.6 \lesssim H/J_3 \lesssim 2.1$. The slight deviation around this region is due to a finite-temperature effect in the Monte Carlo simulations.  In fact, another phase transition occurs at $T/J_3 \sim 0.03$ for $H/J_3=2.0$. 
Thus, most of the low-temperature phases obtained from the finite temperature Monte Carlo simulations shown in Fig.~\ref{Fig:phase-diagram_a8}(a) remain stable down to $T=0$.

\section{Towards more realistic considerations}
\label{sec: more realistic}
So far, we have assumed a periodic array of impurities, which is commensurate with the spin texture. Below, we discuss the stability of the chiral phases upon relaxing this
condition.

\begin{figure}[t!]
  \begin{center}
    \includegraphics[width=0.9 \hsize,bb=0 0 280 194]{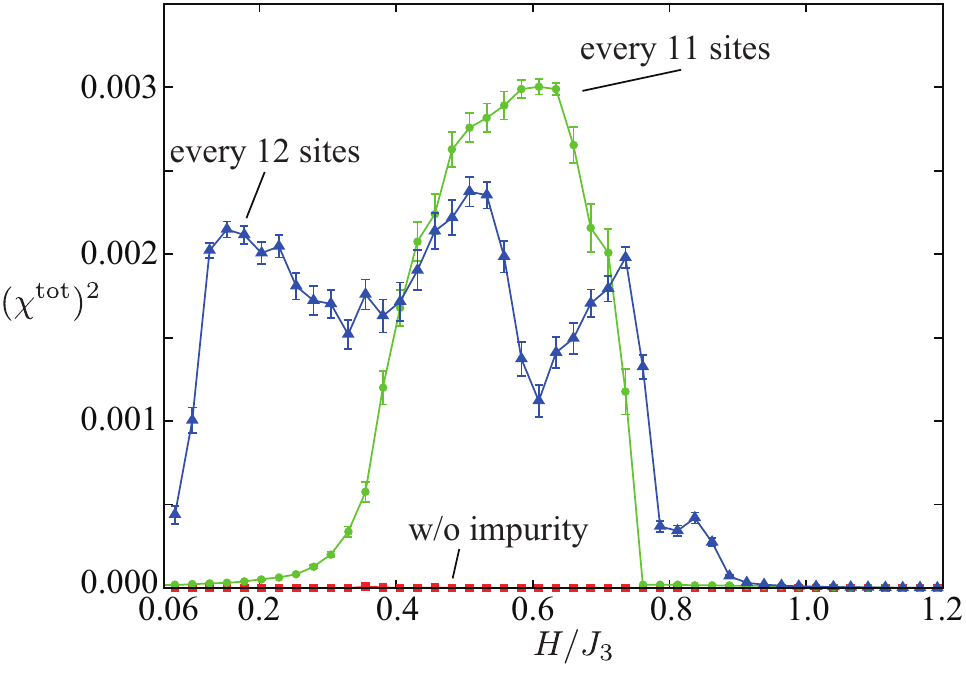} 
    \caption{
      \label{Fig:diffperiodic_chi2}
      $H$-dependence of $(\chi^\text{tot})^2$ for $J_3 / \lvert{J_1}\rvert \approx 0.394$ ($Q = 2\pi/6$) at $T/J_3 \approx 0.1268$ in the model with or without periodic impurities: $a_\text{imp} = \infty$ (no impurities), $a_\text{imp} = 12$, and $a_\text{imp} = 11$.
    }
  \end{center}
\end{figure}

\begin{figure*}
\begin{center}
  \includegraphics[width=0.85 \hsize,bb=0 0 740 252]{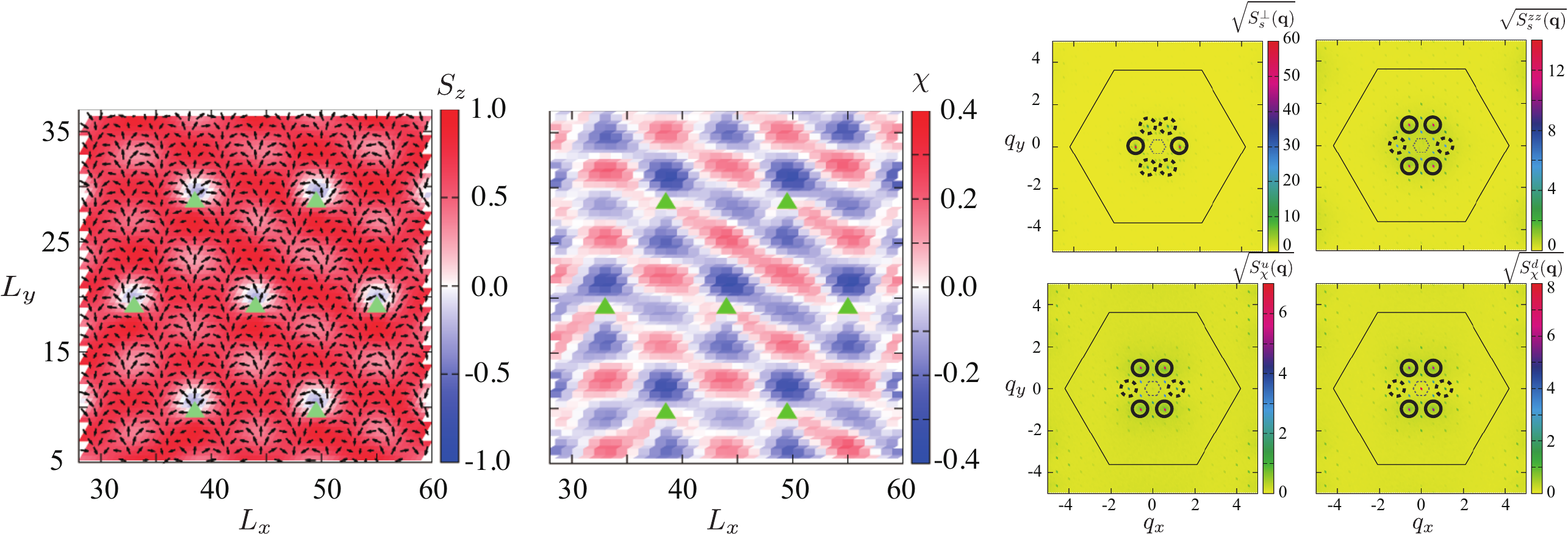}
\caption{
  \label{Fig:a11}
  Snapshots of the spin configuration (left), the chirality texture (middle), and the spin and the chirality structure factors (right) in the model with perfect periodic impurities (triangles) with $a_\text{imp} = 11$ for $J_3 / \lvert{J_1}\rvert \approx 0.394$ ($Q = 2\pi/6$) at $H / J_3 \sim 0.634$.
  The $\mathbf{q}=0$ component is removed from $S_s^{zz}(\mathbf{q})$.
 We average over 500 MCS to integrate out short wavelength fluctuations.
}
\end{center}
\end{figure*}

\subsection{Different impurity concentrations}
\label{sec:Deviation from simple commensurability}

First we investigate the stability of the chiral phases shown in Fig.~\ref{Fig:phase-diagram_a8} upon changing the impurity concentration. 
Our numerical results and the stability analysis  have shown that the  multiple-${\mathbf Q}$ structures arise from the fact that $\mathbf{Q}_\nu$ and $\mathbf{Q}_{\nu'}$ ($\nu \neq \nu'$) are connected by $\mathbf{K}_{\pm}$. Below we demonstrate that the multiple-${\mathbf Q}$ structure is suppressed upon changing the periodicity of impurity array, i.e., upon reducing the commensurability effect. 

Figure~\ref{Fig:diffperiodic_chi2} shows the $H$-dependence of the square of the scalar chirality at a low enough $T$ for several different impurity concentrations and $J_3 / \lvert{J_1}\rvert \approx 0.394$ (corresponding to $Q = 2\pi/6 < Q_c$). In addition to the case without impurities, we show the results for $a_\text{imp} = 11$ and $a_\text{imp} = 12$. The case with $a_\text{imp} = 12$ is similar to the case we discussed in Sec.~\ref{sec:a8}, in the sense that $a_\text{imp} = 12$ is twice as large as $2\pi / Q = 6$ and $\mathbf{K}_{+}+\mathbf{K}_{-} = \mathbf{Q}_1$ holds. In fact, the $H$-dependence of $(\chi^\text{tot})^2$ is qualitatively very similar to the previous case shown in Fig.~\ref{Fig:phase-diagram_a8}(e).
Strictly speaking, the other case with $a_\text{imp} = 11$ is also commensurate with the optimal  magnetic order. However, the minimal period of 
the combined structures is 44 lattice spacings, which is significantly large and thus almost incommensurate. 
Interestingly enough, we find that the net chirality  is finite in the intermediate magnetic field range, although the functional form is rather different.
The spin configuration is triple-$Q^M$ with vortices located around each impurity (Fig.~\ref{Fig:a11}).
The corresponding chirality wave has a multiple-$Q^\chi$ ferrichiral structure with slightly different profiles in $S_\chi^u(\mathbf{q})$ and $S_\chi^d(\mathbf{q})$. 
Our results  suggest that chiral states resulting from nonmagnetic impurities are rather robust against changing the impurity concentration when ${\mathbf Q}$ is small.

\subsection{Skyrmion crystal induced by small  randomness}
\label{subsec:Skyrmion state by slight charge fluctuations}

\begin{figure}[b]
\begin{center}
\includegraphics[width=0.6 \hsize,bb=0 0 685 594]{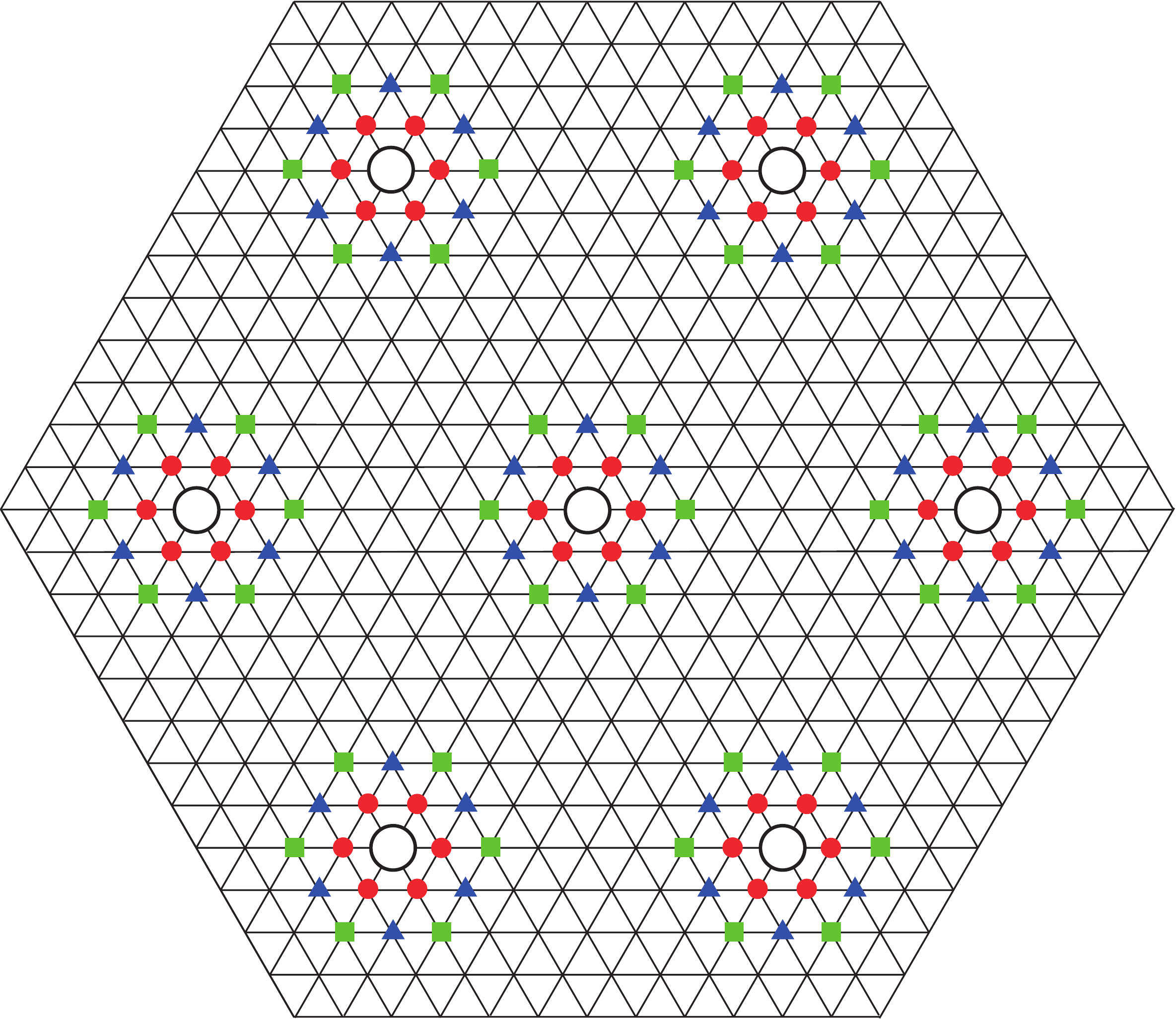} 
\caption{
  \label{Fig:randomness}
  Allowed positions of a nonmagnetic impurity. Partial randomness is included by locating the nonmagnetic impurities either at a regular superlattice site (the open circle at the origin) or at 
  one of its  neighbors (see the text). The filled circles, triangles, and squares denote the NN,  second NN, and the third NN sites, respectively.
   The average distance between closest impurities is $\overline{a}_\text{imp} = 8$.  
}
\end{center}
\end{figure}

\begin{figure}[b]
\begin{center}
\includegraphics[width=0.9 \hsize,bb=0 0 281 241]{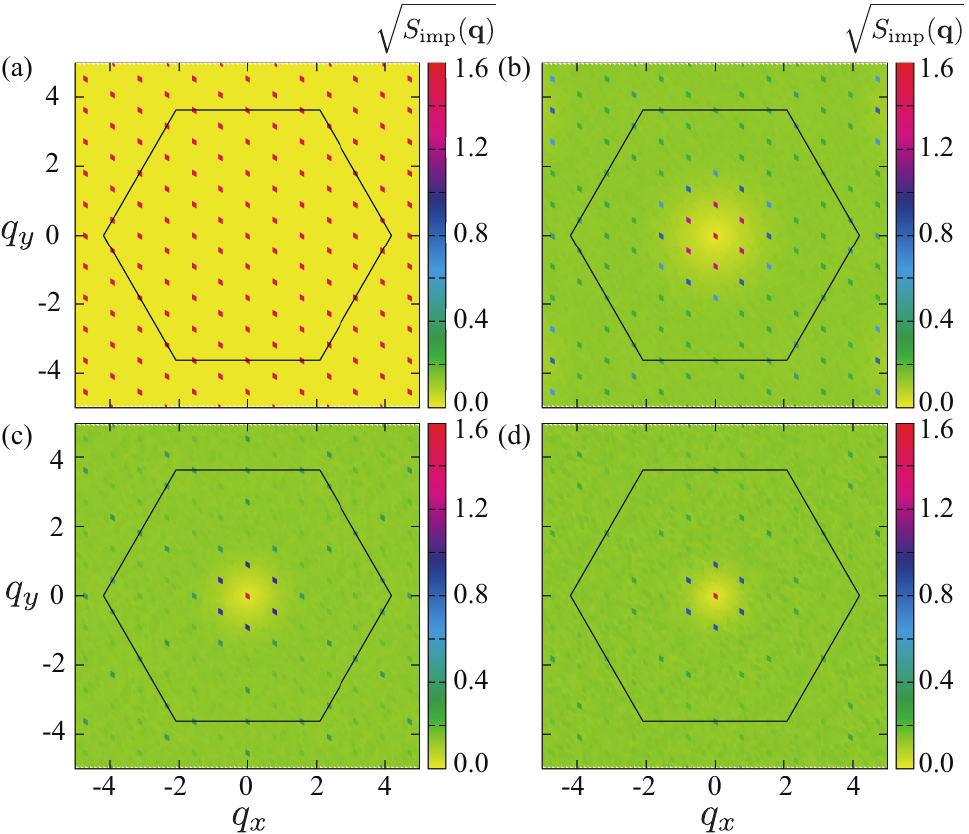} 
\caption{
  \label{Fig:impurity_structure}
    The impurity structure factors (a) for the regular superlattice site and with a randomly distribution up to (b) NN sites, (c) second NN sites, and (d) third NN sites. 
}
\end{center}
\end{figure}

\begin{figure*}
\begin{center}
  \includegraphics[width=0.85 \hsize,bb=0 0 740 252]{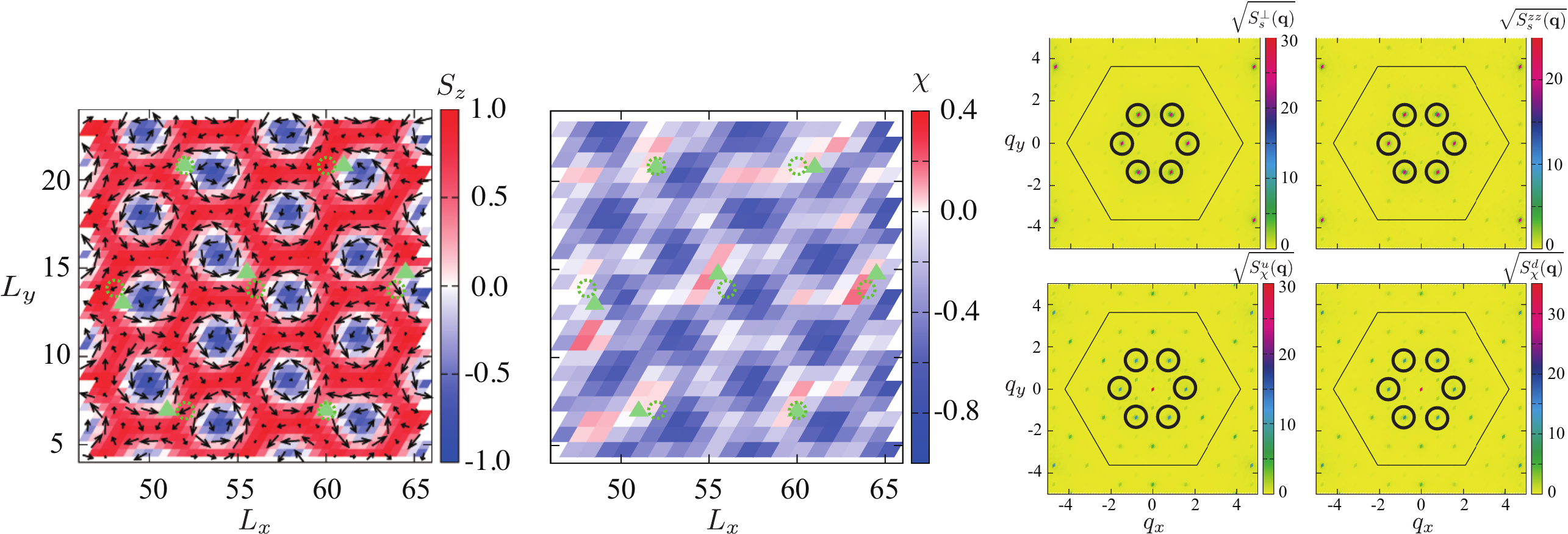}
  \caption{
    \label{Fig:skyrmion}
       Snapshots of the spin configuration (left), the chirality texture (middle), and the spin and the chirality structure factors (right) for periodic impurities \textit{slightly disordered up to NNs} (triangles) with $\overline{a}_\text{imp} = 8$, $H/J_3=1.4$, $T/J_3=0.05$ and $J_3 / \lvert{J_1}\rvert \approx 0.854$ ($Q = 2\pi/4$).
    The dashed circles represent the positions without randomness. 
    The $\mathbf{q}=0$ component has been removed from $S_s^{zz}(\mathbf{q})$.
    We average over 500 MCS to integrate out short wavelength fluctuations.
   }
\end{center}
\end{figure*}

\begin{figure}
\begin{center}
\includegraphics[width=1.0 \hsize,bb=0 0 492 292]{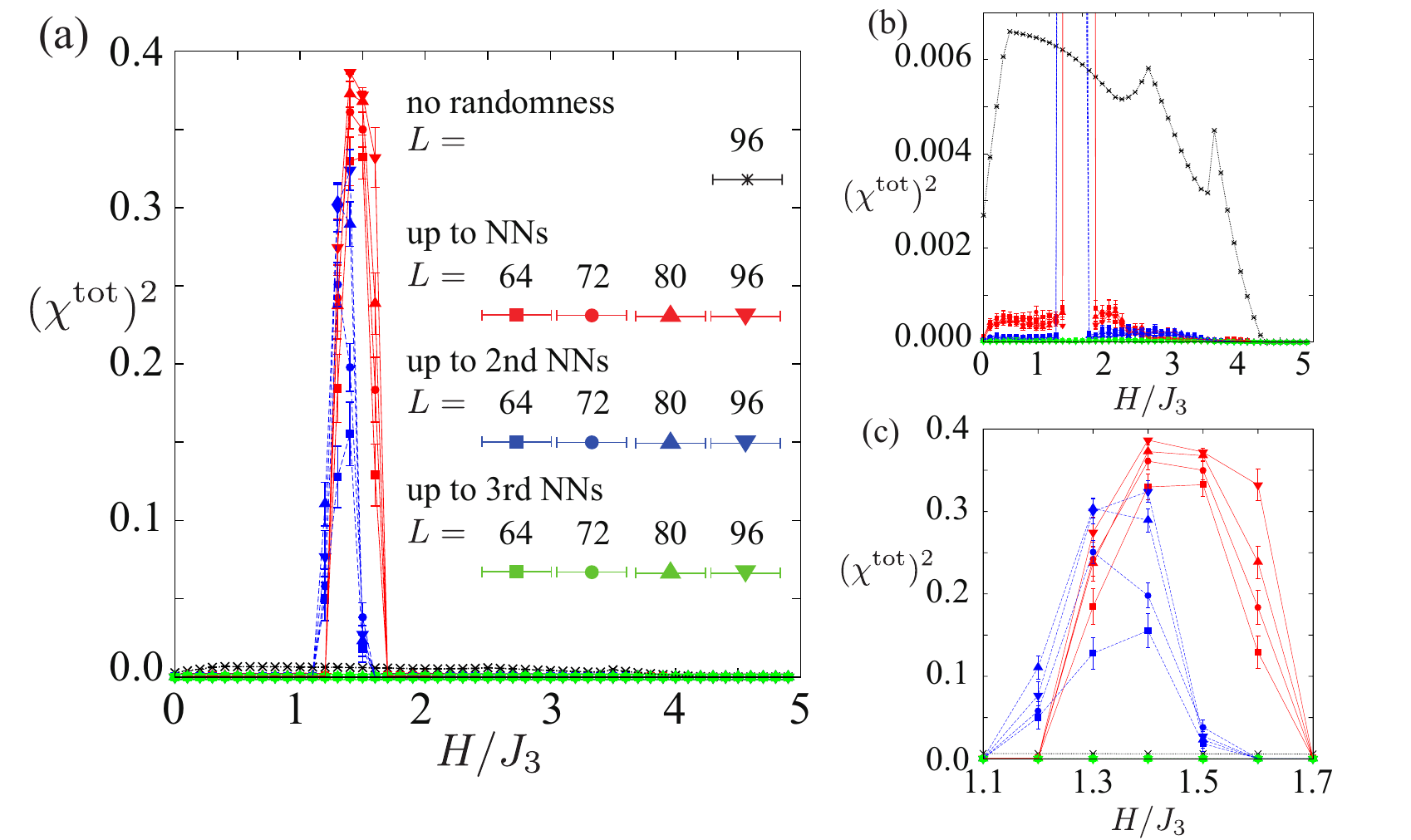} 
\caption{
  \label{Fig:randomness_chi2}
  (a) $H$-dependence of $(\chi^\text{tot})^2$ for the model with  \textit{partial  randomness} of the impurity array [$J_3 / \lvert{J_1}\rvert \approx 0.854$ ($Q = 2\pi/4$) and $T/J_3=0.05$]. (b), (c)  Enlarged views of (a).
}
\end{center}
\end{figure}

Finally, we introduce small quenched randomness into the array of impurities. In the example of a CDW, nonmagnetic ions (holes or doubly-occupied sites) can be frozen at positions slightly away from the perfect superlattice sites. Thus, it is natural to ask what would the consequence of such small quenched randomness in terms of the magnetic ordering. 

To simplify our discussion, we consider the following three cases where, as illustrated in Fig.~\ref{Fig:randomness}, each nonmagnetic ion is uniformly and randomly placed at (i) one of seven sites comprising a site of the perfect superlattice and its six NNs of the underlying lattice, (ii) one of
13 sites comprising a perfect superlattice site, the six NNs, and the six second NNs or (iii) one of 19 sites that include up to the third NN sites. For each case, we take statistical averages by generating $32$--$144$ different impurity configurations.  

First, we show the impurity structure factor
\begin{eqnarray}
  \label{Eq:impstructure}
  S_{\rm imp}(\mathbf{q})=\frac{1}{N} \sum_{j,l} \langle (1-p_j)(1-p_l) \rangle e^{i \mathbf{q}\cdot(\mathbf{r}_j - \mathbf{r}_l)}, 
\end{eqnarray}
for the different randomness (i)--(iii) in Fig.~\ref{Fig:impurity_structure}. In the case of an ideal impurity array, the Bragg peaks appear at $c_1 \mathbf{K}_+ + c_2 \mathbf{K}_-$ ($c_1$ and $c_2$ are integer) with the same intensity, as shown in Fig.~\ref{Fig:impurity_structure}(a). With increasing the degree of randomness from the case (i) in Fig.~\ref{Fig:impurity_structure}(b) through the case (iii) in Fig.~\ref{Fig:impurity_structure}(d), the amplitudes of the Bragg peaks for large $\mathbf{q}$ diminish. In particular, the amplitude at $\mathbf{q} = \mathbf{Q}_\nu$ remains finite (almost disappears) for the cases (i) and (ii) [the case (iii)]. This difference is crucial for the emergence of the skyrmion crystal (see Fig.~\ref{Fig:skyrmion}) that we discuss below. 

As shown in Figs.~\ref{Fig:randomness_chi2}(a)--\ref{Fig:randomness_chi2}(c), we compute $(\chi^\text{tot})^2$ at a low $T$ for $J_3 / \lvert{J_1}\rvert \approx 0.854$ ($Q = 2\pi / 4$). While there is no net 
chirality  in the low and high magnetic field regions, we find that the scalar chirality is drastically enhanced in the intermediate field regime $1.2 \lesssim H / J_3 \lesssim 1.6$ for the moderate impurity randomness, i.e., the cases of (i) and (ii), as shown in Fig.~\ref{Fig:randomness_chi2}(c).
Such a drastic enhancement indicates the emergence of a new phase induced by the quenched randomness. 
The spin configuration corresponds to a triple-$Q^M$ hexagonal skyrmion crystal (Fig.~\ref{Fig:skyrmion}), similar to the state found by Okubo \etal in a different parameter regime of the same model without impurities~\cite{Okubo_PhysRevLett.108.017206}. A similar phase is also obtained for the same model with easy-axis anisotropy.~\cite{leonov2015multiply,Shizeng_PhysRevB.93.064430,Hayami_PhysRevB.93.184413} 
We  note that, unlike the ferrochiral 3$Q^M$ spiral state in Fig.~\ref{Fig:phase5-8_a8}(f), the chirality structure factor of this state shows six peaks at $\mathbf{q}=\pm\mathbf{Q}_{1\le\nu\le3}$, in addition to the $\mathbf{q}=\mathbf{0}$  peak  (see also Appendix~\ref{sec:Finite-size scaling of each phase}).

The local spin reorientation induced by impurities is central to explain why the skyrmion crystal appears only for small charge randomness in the range of $1.2 \lesssim H /J_3 \lesssim 1.6$.
Without the randomness, the ferrichiral 3$Q^M$-2$Q^\chi$ spiral I phase is realized in this region, where the spin texture creates the vortex configuration with $S^z \approx 0$ around each impurity.
This state is more stable than the skyrmion crystal in the absence of randomness, because impurities would be near the center of the skyrmions (commensurability effect). Such a situation is energetically unfavorable due to the large $|S^z|$ value of the spins near the skyrmion core.
(According to our preliminary considerations, the energy is minimized by increasing the $xy$ components of the spins near the nonmagnetic impurities relatively to the other spins.)
By introducing the small quenched randomness, as shown in the left panel of Fig.~\ref{Fig:skyrmion}, the impurities can escape from the skyrmion core towards the perimeter region where $S^z \approx 0$ and nucleate antivortices around themselves. 
While the chirality structure is to a large extent characterized by the $\mathbf{q} = 0$ component, as shown in the middle panel in Fig.~\ref{Fig:skyrmion}, there are spots with the opposite sign around the impurities because of the induced antivortices. In contrast, the randomness increases the energy of the ferrichiral 3$Q^M$-2$Q^\chi$ spiral I phase because it  pushes the impurities away from their ``comfortable'' $S^z \approx 0$ zone. 
This is also the reason why the skyrmion phase is destabilized when the randomness becomes too strong, i.e., in the case (iii) (see Fig.~\ref{Fig:randomness_chi2}). Once the typical deviation exceeds the skyrmion radius, the impurities are pushed into the large $|S^z|$ region again.
This result points to a new mechanism of stabilizing skyrmion crystals, namely, moderate randomness in the array of nonmagnetic impurities can induce a chiral phase when combined with frustrated exchange interactions.

\section{Summary and discussion}
\label{sec: summary}

As we discussed above, CDW ordering of a strongly-coupled single-band model away from half-filling can provide a natural realization of
our periodic array of non-magnetic impurities. Moreover, it is natural to expect that charge orderings which are commensurate with the 
magnetic ordering wave-vectors will be naturally selected at the corresponding filling fractions of the Hubbard model. Our previous analysis of 
the effect of small randomness in the periodic array of impurities suggests the interesting possibility of having skyrmion crystals stabilized by the zero 
point fluctuations of the CDW. 

What are the alternative realizations of the model studied in this work? Here we present three additional proposals for realizing periodic arrays of nonmagnetic impurities.
The first realization involves surface science technology.~\cite{eigler1990positioning}  For instance, a selective atom substitution based on the scanning tunneling microscope technique enables  manipulate of atoms on the surface of Mott insulators with  spiral order, such as Fe$_x$Ni$_{1-x}$Br$_2$~\cite{moore1985magnetic} and Zn$_x$Ni$_{1-x}$Br$_2$.~\cite{day1981neutron}
The second possibility is  through Kondo lattice systems with long-range Coulomb interaction between conduction electrons. 
Even not taking into account Coulomb interaction explicitly (i.e., in the usual Kondo lattice model) charge ordering can be induced by magnetic ordering~\cite{Sahinur_PhysRevB.91.140403,Misawa_PhysRevLett.110.246401,Hayami:PhysRevB.89.085124,hayami2013charge,ishizuka2012magnetic} and produce a periodic potential for the spin degrees of freedom similar to the one induced by the periodic array of nonmagnetic impurities. Coulomb interactions can enhance this tendency  producing an even stronger CDW ordering. 
The third proposal is based on selective Kondo screening: some heavy fermion compounds are known to exhibit partially ordered magnetic states.
For instance, the partially-disordered compound, UNi$_4$B, exhibits a magnetic vortex  structure~\cite{Mentink1994,Oyamada2007,Hayami_1742-6596-592-1-012101}. A possible scenario is that the partial disorder is produced by a site-selective formation of Kondo singlet states.~\cite{Lacroix,Motome2010,Hayami2011}
The site-selective Kondo screening is then an alternative mechanism for producing a nonmagnetic superlattice.

To summarize, by taking the classical $J_1$-$J_3$ Heisenberg model on the triangular lattice as an example, we have shown that exotic multiple-$\mathbf{Q}$ states can be induced by  periodically distributed nonmagnetic impurities. 
The interplay between the spin configuration and the  underlying impurity superlattice renders most of the states chiral (i.e., with net scalar chirality). We have also shown that weak randomness in the impurity positions, relative to the periodic array, induces a skyrmion crystal phase for intermediate magnetic field values. Our results suggest that a variety of magnetically ordered states with  nonzero net scalar chirality can be realized by changing the concentration of nonmagnetic impurities, magnetic field, and temperature.

\begin{acknowledgments}
Computer resources for numerical calculations were supported by the Institutional Computing Program at LANL. This work was carried out under the auspices of the U.S. DOE contract No. DE-AC52-06NA25396 through the LDRD program. Y.K.~acknowledges the financial supports from the RIKEN iTHES project.  
\end{acknowledgments}

\appendix
\section{Characterization of each phase}
\label{sec:Finite-size scaling of each phase}
\begin{figure*}[p!]
\begin{center}
\includegraphics[width=0.75 \hsize]{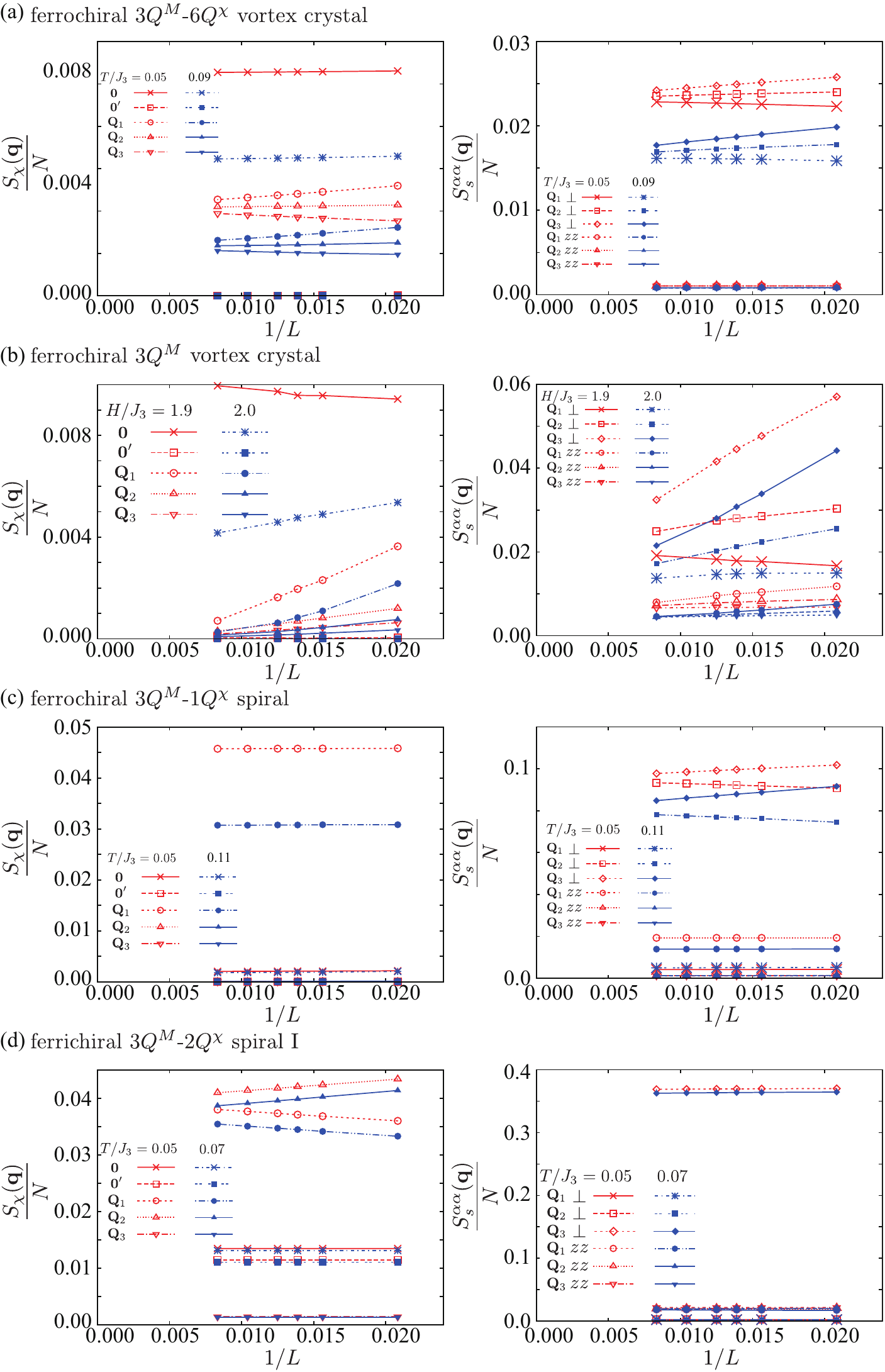}
\caption{
\label{Fig:sizedep_opt1-4}
Size dependence of the order parameters for each phase described in the phase diagram of Fig.~\ref{Fig:phase-diagram_a8}(c): the chirality structure factor evaluated at $\mathbf{q}=\mathbf{0}$, $\mathbf{0}'$, $\mathbf{Q}_1$, $\mathbf{Q}_2$, $\mathbf{Q}_3$ [$S_{\chi}(\mathbf{q})=S_{\chi}^{u}(\mathbf{q})+S_{\chi}^{d}(\mathbf{q})$ for $\mathbf{q}=\mathbf{0}$, $\mathbf{Q}_1$, $\mathbf{Q}_2$, and $\mathbf{Q}_3$ and $S_{\chi}(\mathbf{0}')=S_{\chi}^{u}(\mathbf{0})-S_{\chi}^{d}(\mathbf{0})$] and the magnetic structure factor at $\mathbf{q}=\mathbf{Q}_1$, $\mathbf{Q}_2$, $\mathbf{Q}_3$ in (a) the ferrochiral 3$Q^M$-6$Q^\chi$ vortex crystal phase ($H/J_3=3.8$), (b) ferrochiral 3$Q^M$ vortex crystal phase ($T/J_3=0.41$), (c) ferrochiral 3$Q^M$-1$Q^\chi$ spiral phase for $Q=2\pi/4$ ($H/J_3=3.0$), and (d) ferrichiral 3$Q^M$-2$Q^\chi$ spiral I phase ($H/J_3=1.0$) for $Q=2\pi/4$ and different temperature and magnetic fields. 
}
\end{center}
\end{figure*}

\begin{figure*}[p!]
\begin{center}
\includegraphics[width=0.75 \hsize]{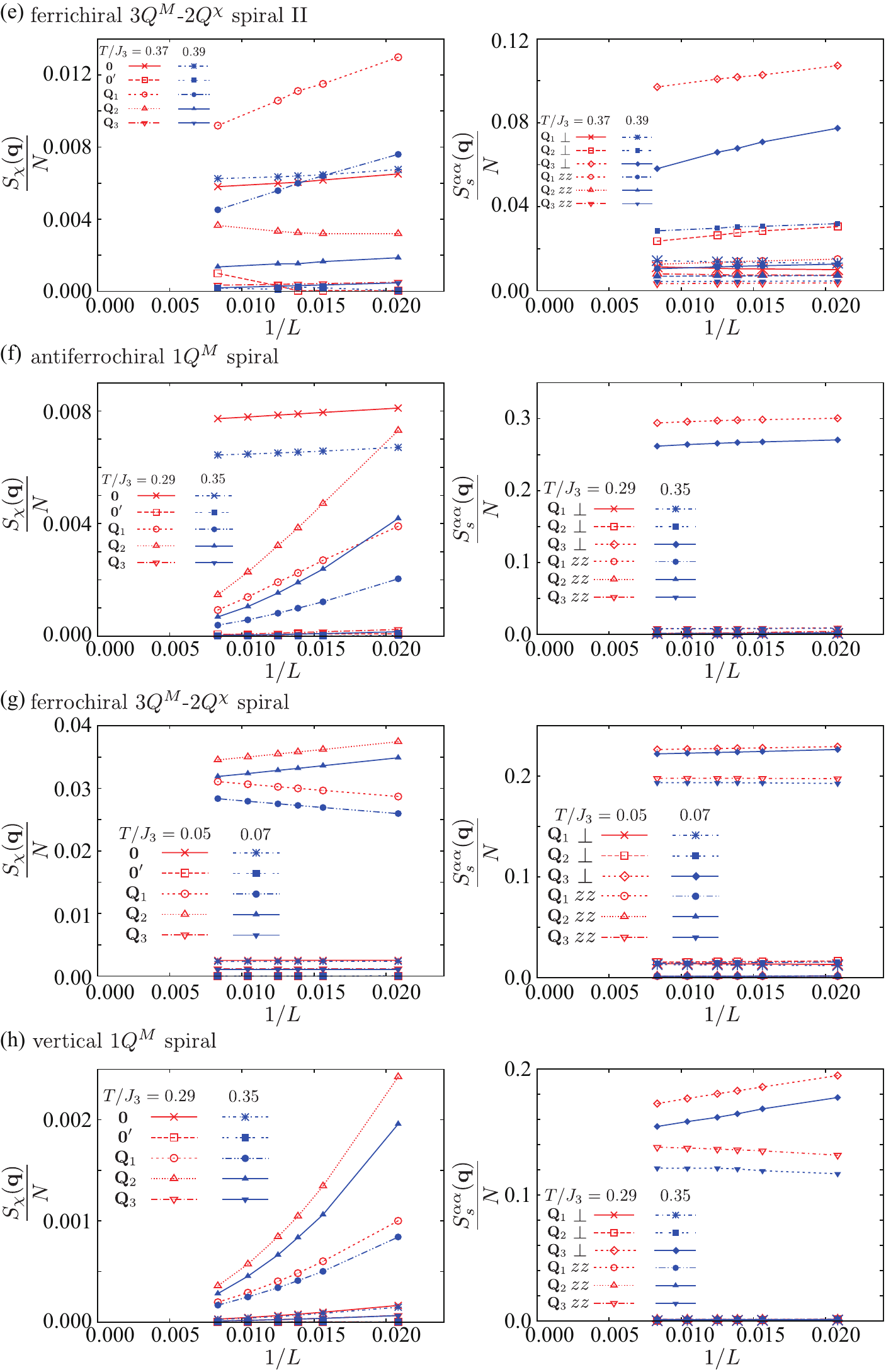} 
\caption{
\label{Fig:sizedep_opt5-8}
Size dependence of the order parameters for each phase described in the phase diagram of Fig.~\ref{Fig:phase-diagram_a8}(c): the chirality structure factor evaluated at $\mathbf{q}=\mathbf{0}$, $\mathbf{0}'$, $\mathbf{Q}_1$, $\mathbf{Q}_2$, $\mathbf{Q}_3$ [$S_{\chi}(\mathbf{q})=S_{\chi}^{u}(\mathbf{q})+S_{\chi}^{d}(\mathbf{q})$ for $\mathbf{q}=\mathbf{0}$, $\mathbf{Q}_1$, $\mathbf{Q}_2$, and $\mathbf{Q}_3$ and $S_{\chi}(\mathbf{0}')=S_{\chi}^{u}(\mathbf{0})-S_{\chi}^{d}(\mathbf{0})$] and the magnetic structure factor at $\mathbf{q}=\mathbf{Q}_1$, $\mathbf{Q}_2$, $\mathbf{Q}_3$ in (e) ferrichiral 3$Q^M$-2$Q^\chi$ spiral II phase ($H/J_3=2.0$), (f) antiferrochiral single-$Q^M$ spiral phase ($H/J_3=1.0$), (g) ferrochiral 3$Q^M$-2$Q^\chi$ spiral phase ($H/J_3=0.2$), and (h) vertical single-$Q^M$ spiral phase ($H/J_3=0.2$) for $Q=2\pi/4$ and different temperature and magnetic fields. 
}
\end{center}
\end{figure*}

\begin{figure*}[htb!]
\begin{center}
\includegraphics[width=0.75 \hsize]{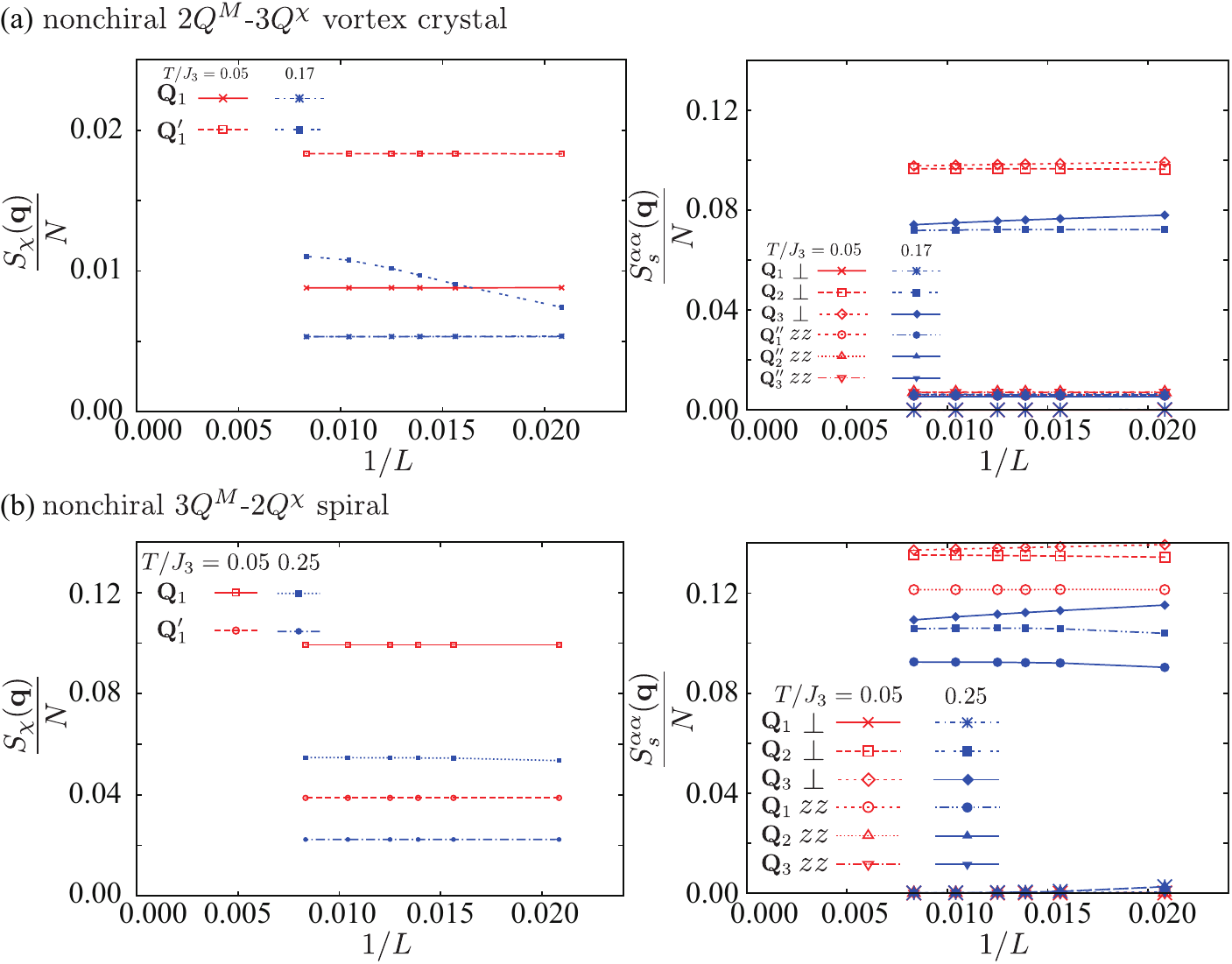} 
\caption{
\label{Fig:sizedep_opt_2times}
Size dependence of the order parameters for two phases described in the phase diagram of Fig.~\ref{Fig:phase-diagram_a4}(b): (a) the $xy$- and $z$-components of the spin structure factor evaluated at $\mathbf{q}=\mathbf{Q}_1$, $\mathbf{Q}_2$, $\mathbf{Q}_3$ and the chirality structure factor evaluated at $\mathbf{q}=\mathbf{Q}_1$ and $\mathbf{Q}_1'$ [$Q_1' = \pi/(2\sqrt{3})$] in the nonchiral 3$Q^M$-2$Q^\chi$ spiral phase ($H/J_3=1.0$) and (b) the $z$($xy$)-component of the spin structure factor evaluated at $\mathbf{q}=\mathbf{Q}_1''$, $\mathbf{Q}_2''$, and $\mathbf{Q}_3''$ where $Q_\nu'' = \pi/\sqrt{3}$ ($\mathbf{q}=\mathbf{Q}_1$, $\mathbf{Q}_2$, and $\mathbf{Q}_3$) and the chirality structure factor evaluated at $\mathbf{q}=\mathbf{Q}_1$ and $\mathbf{Q}_1'$ in the nonchiral 2$Q^M$-3$Q^\chi$ vortex crystal phase  ($H/J_3=3.0$). 
}
\end{center}
\end{figure*}

Figures~\ref{Fig:sizedep_opt1-4}, \ref{Fig:sizedep_opt5-8}, and  \ref{Fig:sizedep_opt_2times} show the $1/L$ dependence of the $xy$ and $z$ components of the spin structure factor and the chirality structure factor normalized by the system size $N$.
They should scale $S(\mathbf{q})  \sim \bigO{1}$, $L^{2-\eta}$, and $N$, respectively, when the corresponding mode is disordered, critical, and long-range ordered. Note that the Mermin-Wagner theorem precludes the long-range order in the $xy$ component at finite $T$. 

%\bibliographystyle{apsrev}
%\bibliography{ref}

\nocite{apsrev41Control}
\bibliographystyle{my-apsrev4-1}
\bibliography{my-refcontrol,ref}

\end{document}